\documentclass[a4paper,11pt]{article}
\usepackage[utf8]{inputenc}
\usepackage{lmodern}
\usepackage[T1]{fontenc}
\usepackage{amsmath,amsthm,amsfonts,amssymb,bm}
\usepackage{dsfont}			
\usepackage{enumitem}
\usepackage{graphicx}
\usepackage{booktabs}
\usepackage{xcolor}
\usepackage{authblk}
\usepackage[authoryear]{natbib}
\usepackage[right=2.5cm,left=2.5cm,top=2cm,bottom=3cm]{geometry}
\usepackage[font={small,it}]{caption}	
\usepackage[colorlinks,linkcolor=blue,citecolor=blue,urlcolor=blue]{hyperref}
\usepackage{subfig}
\usepackage{mathtools}

\usepackage{multirow}
\usepackage{makecell}
\usepackage{afterpage}

\title{ \textsc{Parsimonious Bayesian Factor Analysis for modelling latent structures in spectroscopy data} }
\author[1,3]{Alessandro Casa}
\author[2,3]{Tom F. O'Callaghan}
\author[1,3]{Thomas Brendan Murphy}
\affil[1]{School of Mathematics \& Statistics, University College Dublin}
\affil[2]{School of Food \& Nutritional Sciences, University College Cork}
\affil[3]{Vistamilk SFI Research Centre}
\date{}                     

\begin{document}

\maketitle

\begin{abstract}
In recent years, within the dairy sector, animal diet and management practices have been receiving increased attention, in particular examining the impact of pasture-based feeding strategies on the composition and quality of milk and dairy products, in line with the increased prevalence of premium \emph{grass-fed} dairy products appearing on market shelves. To date, there are limited testing methods available for the verification of \emph{grass-fed} dairy and as a consequence these products are susceptible to food fraud and adulteration. Therefore, with this in mind, enhanced statistical tools studying potential differences among milk samples coming from animals on different feeding systems are required, thus providing increased security around the authenticity of the products. 

Infrared spectroscopy techniques are widely used to collect data on milk samples and to predict milk related traits and characteristics. While these data are routinely used to predict the composition of the macro components of milk, each spectrum also provides a reservoir of unharnessed information about the sample. The accumulation and subsequent interpretation of these data present a number of challenges due to their high-dimensionality and the relationships amongst the spectral variables. 

In this work, directly motivated by a dairy application, we propose a modification of the standard factor analysis to induce a parsimonious summary of spectroscopic data. The proposed procedure maps the observations into a low-dimensional latent space while simultaneously clustering observed variables. The proposed method indicates possible redundancies in the data and it helps disentangle the complex relationships among the wavelengths. A flexible Bayesian estimation procedure is proposed for model fitting, providing reasonable values for the number of latent factors and clusters. The method is applied on milk mid-infrared (MIR) spectroscopy data from dairy cows on distinctly different pasture and non-pasture based diets, providing accurate modelling of the data correlation, the clustering of variables in the data and information on differences between milk samples from cows on different diets. 
\end{abstract}


\section{Introduction}
In recent years, the food industry has gone through rapid changes, partially due to ever evolving consumer preferences and increased consumers' awareness of food and health. We are currently at the forefront of growing demand for detailed and accurate knowledge concerning food quality, authentication and security. The sectors potentially most vulnerable to these changing trends are those processing and preparing foodstuffs of animal origin. As a consequence, the dairy sector has been particularly involved in this transition, with an increasing attention towards product quality, traceability and adherence to procedures respectful to animal welfare and the environment. 

In this scenario, one of the aspects which is gaining more attention and relevance is concerned with cattle feeding regimen. In general consumers regard pasture based feeding as more respectful of the animal well-being, producing products that are more natural and healthy \citep{elgersma2012new}. These perceptions appear to have some basis in fact as mounting evidence has demonstrated that outdoor pasture based feeding of cows results in milk and dairy products with enhanced beneficial nutrients compared to indoor total mixed ration (TMR) like systems \citep{ocall2016effect,o2016quality,o2017effect}. Furthermore, pasture based feeding has been demonstrated to lead to ameliorations in the organolectic characteristics of dairy products providing signature like traits \citep{faulkner2018,alothman2019grass,c2020cross} and improved quality. As such, in many markets, \emph{grass-fed} dairy products usually demand a premium price from consumers, and as often happens with expensive products, milk produced by grass fed cows is susceptible to food adulteration and fraud. As a direct consequence, there has been an increased requirement for development of methods capable to detect the presence of adulterants and authenticate the traceability of the milk \cite[see][for a recent review]{kamal2015analytical}. In the literature several different approaches which have been proposed rely on the identification of one or more useful biomarkers in order to build meaningful authentication methods \cite[see][and references therein]{capuano2014verification}. Furthermore, \cite{ocall2016effect,o2018pasture} highlighted the ability of fatty acid profiling coupled with multivariate analysis and H-NMR to be able to distinguish between milk from pasture and TMR based diets. Nonetheless, these techniques are considered to be expensive and time consuming since they require laboratory extraction routines to collect the data, compromising their widespread effective utility. 

\begin{figure}[t]
  \centering 
  \includegraphics[height = 8cm, width = 13cm]{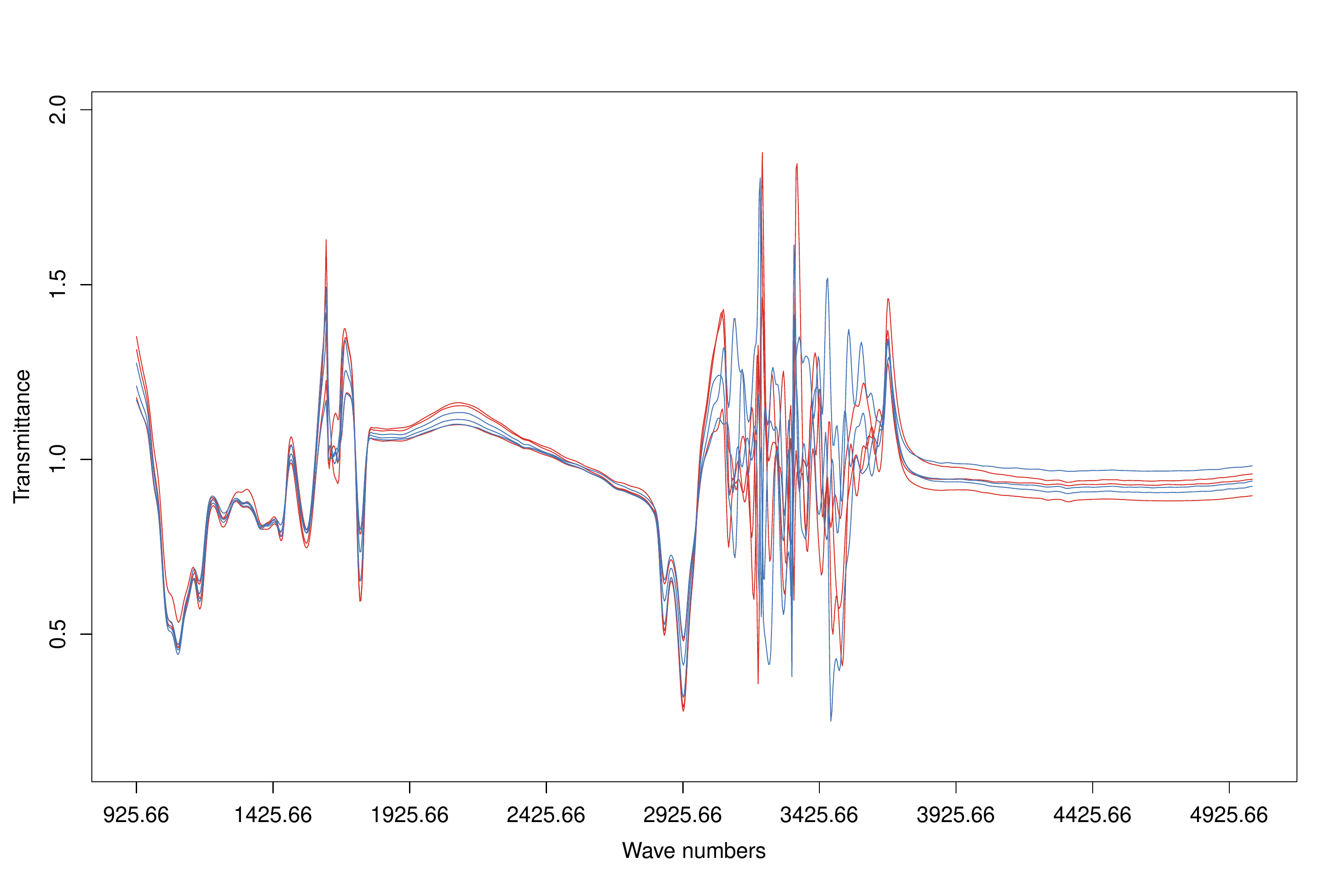}
  \caption{The mid-infrared spectra recorded for a subset of the analyzed milk samples corresponding to different diet regimens. The samples are colored as pasture-diet = red, total mixed ration-diet = blue.}
  \label{fig:fig_example_waves}
\end{figure}

On the other hand vibrational spectroscopy techniques, such as Fourier transform near-infrared (NIR) and mid-infrared (MIR) spectroscopy, are known to be cheap, rapid and non-disruptive alternatives to collect large amounts of data and to analyze different biological materials. Such methods have been already proven useful in food authenticity studies \citep[eg.][]{murphy2010}; readers may refer to \citet{downey1996authentication} and \citet{reid2006recent} for more thorough reviews. When a material is analyzed via MIR spectroscopy the light is passed through a sample of that material at a sequence of wavelengths in the mid-infrared region (900 to 5000 cm$^{-1}$). The passage of the light activates the sample's chemical bonds leading to an absorption of energy from the light itself. The amount of energy absorbed or transmitted by the sample, at different wavelengths, creates the spectrum of that sample that might be subsequently used to analyze its characteristics (see Figure \ref{fig:fig_example_waves} for a graphical illustration of some MIR spectra). 

Vibrational spectroscopy techniques have already been fruitfully used in the dairy framework to determine milk characteristics such as protein, lactose and casein concentration \citep{de2014invited} as well as to predict milk and animal related traits such as milk fatty acids \citep{bonfatti2017comparison} and energy efficiency and intake \citep{mcparland2014mid,mcparland2016potential}. However, the usefulness of infrared spectroscopy data to authenticate cow feeding regimens has been less widely explored even though standard classification tools have produced promising results in \citet{coppa2012authentication} and \citet{valenti2013infrared}.

\begin{figure}[ht]
\centering
\begin{minipage}{.47\textwidth}
  \centering
  \includegraphics[height = 6cm, width=6cm]{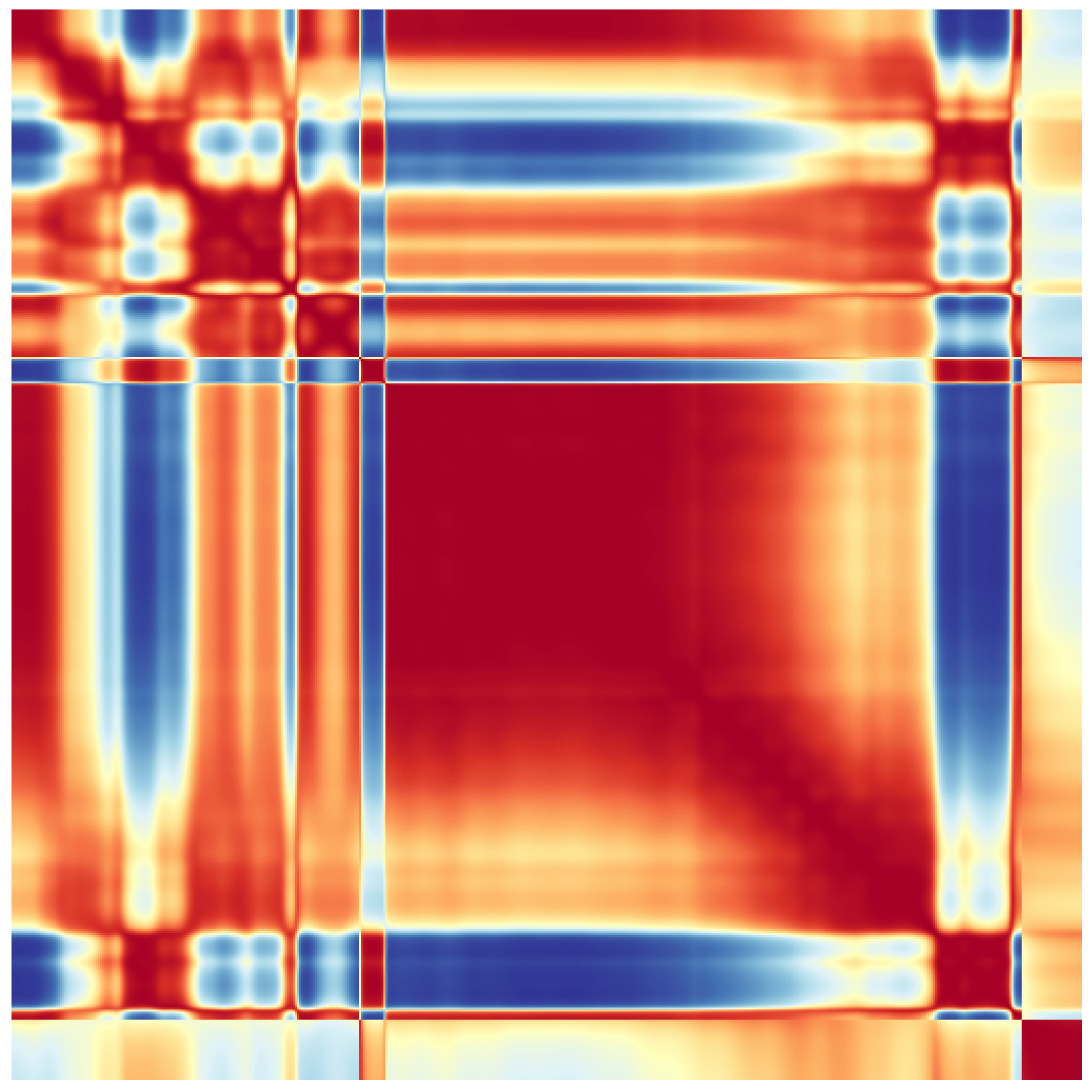}
\end{minipage}%
\begin{minipage}{.47\textwidth}
  \centering
  \includegraphics[height = 6cm,width = 6.5cm]{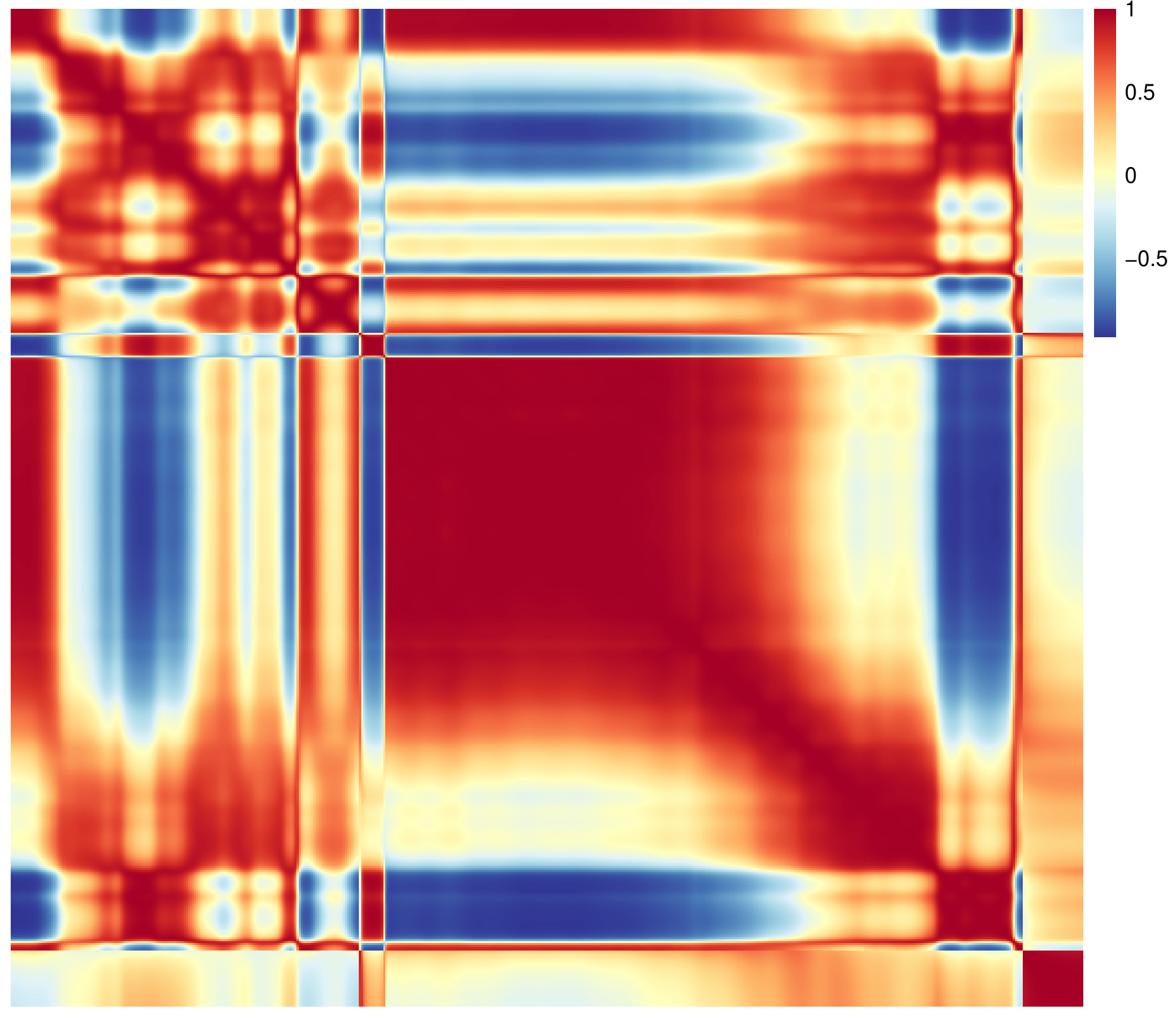}
\end{minipage}
  \caption{Sample correlation matrices computed on the milk samples produced by pasture fed cows (on the left) and total mixed ration fed cows (on the right).}
 \label{fig:sampleCor}
\end{figure}

Therefore, a thorough exploration of the features of spectroscopy data when used to analyze samples coming from differently fed animals is somehow still missing. From a statistical point of view NIR and MIR spectroscopy data introduce some challenges that have to be carefully addressed. The first one is concerned with their high-dimensionality since, usually, each single observation consists of more than 1000 transmittance or absorbance values over the MIR or NIR regions. Moreover these data are highly correlated with underlying chemical processes entailing rather complex correlation structures, as can be seen in Figure \ref{fig:sampleCor}. From the figure it is clear how, albeit adjacent wavelengths tend to be highly correlated, strong relationships are witnessed among regions of the spectrum far from each other. The information contained in each spectrum is indeed known to be structured in a rather complicated way and possibly spread over different locations. 

For these reasons, statistical methodologies being able to provide a parsimonious representation of the correlation structures in spectroscopy data turn out to be particularly useful. On one hand they can mitigate high-dimensionality related issues, by summarizing parsimoniously the information in a lower dimensional representation. On the other hand a proper reconstruction of the relations seen in Figure \ref{fig:sampleCor} may help in identifying which wavelength regions are carrying similar information, when the aim is to discriminate between samples coming from differently fed animals. This identification, when coupled with subject-matter knowledge, can highlight which chemical structures are responsible for the structures seen in the milk sample spectra. Moreover, it can be useful to identify the main nutritional differences implied in the milk from different diet regimens, serving as a stepping stone for classification purposes.

When analyzing spectroscopy data latent variable models are considered to be the state of the art to tackle some of the challenges mentioned above. Techniques such as \emph{Partial Least Squares} (PLS) and \emph{principal component analysis} (PCA) are widely used both for predictive purposes and to reduce the dimensionality of the data, by summarizing the information in a lower number of newly built features. In a similar fashion, \emph{factor analysis} \citep[FA,][]{everitt1984,bartholomew2011latent} provides a parsimonious representation of the observed data, by building new variables called \emph{factors}, while simultaneously explaining the correlation among high-dimensional observations. For this reason, when the aim is to reconstruct structures as the ones in Figure \ref{fig:sampleCor}, factor analysis represents a suitable strategy to follow. Nonetheless, even if FA effectively reduces the dimensionality of the data, standard FA does not provide information about possible redundancies in the observed features. For this reason, in this work, we propose a suitable modification of the standard factor analysis model which allows the detection of redundant variables and which produces a partition of the variables themselves, thus possibly gaining useful insights about similarly behaving spectral regions. 

The rest of the paper is structured as follows. In Section \ref{sec:data_description} we described the mid-infrared spectroscopy data which motivates our proposal. In Section \ref{sec:model_formulation} we outline the proposed methodology with a specific focus on the proposed Bayesian estimation procedure and on the involved model selection steps. Some analyses on synthetic datasets are reported in Section \ref{sec:simulations}, while in Section \ref{sec:application} we present the results  obtained on the milk spectroscopy data. Finally, in Section \ref{sec:finalremarks}, we conclude with some final remarks and highlight some advantageous avenues for future research. 

\section{Dairy diet MIR spectroscopy data}\label{sec:data_description}
The data we consider in this study consists of a collection of mid-infrared spectra from 4320 milk samples produced from cattle on three dietary treatments over a three year period on the Teagasc Moorepark Dairy research Farm (Fermoy, Co. Cork, Ireland). The data are comprised of spectra extracted from morning (am) and evening (pm) milk samples collected weekly from Holstein-Freisian cows using a ProFoss FT6000 series instruments (FOSS, Ireland) between the period of May and August in 2015, 2016 and 2017. In each year 54 cows were randomly assigned to each of the dietary treatment for the entire lactation period of that year. Treatments included grass (GRS) which consisted of cows maintained outdoors on a perennial ryegrass sward only, clover (CLV) whereby cows were maintained outdoors on a perennial ryegrass with 20\% white clover sward only and total mixed ration (TMR) where cows were maintained indoors year round and nutrients are combined together in a single nutritional mix consisting of grass silage, maize silage and concentrates. For further information on the experimental design and dietary treatments, see \citet{faulkner2018}, \cite{o2016quality}, \citet{ocall2016effect}, \citet{o2017effect} and \citet{o2018pasture}. More specifically, 2931 samples come from cows being fed with pasture (GRS and CLV), while the remaining 1389 come from cows being fed with TMR. Note that the original milk samples may be grouped according to the feeding system into three different classes (GRS, CLV and TMR) rather than two (pasture and TMR). In practice, in our work the first two classes have been merged together into a general pasture-based diet group, because of their strong similarities from a compositional perspective. 

The total number of cows involved in the experiment is equal to 120, thus implying that multiple animal measurements are available with a mean number of 36 samples per cow. The samples have been collected following a yearly balanced scheme and they represent a balance of different parities. The samples considered in this work have been restricted to the ones being collected mainly in the summer months since it represents a period of milk production with highest prevalence of grass growth. 
Note that, for each sample, a spectrum consists of 1060 transmittance measurements in the region going from 925 cm$^{-1}$ to 5010 cm$^{-1}$. Finally, we have some additional information about fat, protein and lactose content in the available milk samples obtained using channels on the FT6000 calibrated against wet chemistry results. 

\section{Factor analysis with redundant variables}\label{sec:model_formulation}
\subsection{Framework and model specification}\label{sec:model_definition} 
As mentioned in the introduction standard Factor Analysis (denoted as FA in the following) provides a convenient and parsimonious representation of the dependence structure among high-dimensional observations by mapping them in a low-dimensional latent space.  

Let $X = \{x_1, \dots, x_n \}$, with $x_i \in \mathbb{R}^p$, the set of the observed data. Factor analysis models each observation $x_i$ as a linear combination of latent variables, called factors, as follows 
\begin{equation}\label{eq:stdFA}
x_i = \mu + \Lambda u_i + \varepsilon_i, \hspace*{1cm} i = 1, \dots, n,
\end{equation}
where $\mu \in \mathbb{R}^p$ is the mean vector, $\Lambda = \{ \lambda_{jk}\}_{j=1,\dots,p , k=1,\dots,K}$ is a $p \times K$ factor loadings matrix with $K$ being the number of factors, $u_i \in \mathbb{R}^K$ denotes the factor scores vector while $\varepsilon_i \sim \mathcal{N}_p(0, \Psi)$ is an idiosyncratic error term where $\Psi = \text{diag}(\psi_1,\dots,\psi_p)$, with $\psi_j$'s often referred to as the uniquenesses. Without loss of generality we assume in the following that the data have been centered, hence $\mu = 0$. Moreover, the latent factors are assumed to be normally distributed with zero mean and covariance equal to the identity matrix. \\
Consequently, we have that $(x_i|u_i) \sim \mathcal{N}_p(\Lambda u_i, \Psi)$. On the other hand marginally $x_i$ is distributed according to a Gaussian distribution with zero mean and covariance matrix 
\begin{equation}\label{eq:covFA}
\Sigma = \Lambda\Lambda^T + \Psi \; .
\end{equation}
In practical applications the number of variables $p$ is considerably higher than the number of factors $K$. Therefore the decomposition in (\ref{eq:covFA}) introduces a convenient and parsimonious representation of the relationships among the observed features in high-dimensional settings. 

From (\ref{eq:covFA}), and recalling that $\Psi$ is a diagonal matrix, it follows, in standard FA, that the correlation between the original variables is modelled via the loading matrix $\Lambda$. Therefore, in the literature, the attention has been focused on how to model and estimate $\Lambda$ appropriately. In recent years a lot of different solutions have been proposed, both from a frequentist \citep[see e.g.][]{hirose2012variable,hirose2015sparse} and from a Bayesian \citep[see e.g][]{bhattacharya2011sparse,legramanti2020bayesian} standpoint, to obtain sparse estimates of the loading matrix. Setting some values of $\Lambda$ exactly equal to zero allows an even more parsimonious representation of the original covariance structure. Moreover, it could be convenient from an interpretative point of view since relating each factor to a smaller number of observed variables helps to give meaning to the factors themselves. Lastly, note that if all the elements in a row of the loading matrix are equal to zero we obtain an indication about the uninformative nature of the $j$-th variable, being uncorrelated with all the other features and essentially represented as noise. 

The detection of uninformative features in the FA framework is tackled in the aforementioned references, but to the best of our knowledge possible redundancy has not been tackled yet. A variable is defined as redundant when it carries information similar to the one provided by another variable (or variables), usually due to the strong correlation between them. The effective detection of redundancies, which is a challenging task, can lead to more parsimonious modelling. Note that, as briefly introduced in the previous sections, and as it is graphically represented in Figure \ref{fig:sampleCor}, redundancy can be a complex issue when analyzing spectroscopy data. Therefore, in order to properly account for it, in this work we introduce a model in which some of the variables are mapped into the latent space by means of the same loading coefficients thus giving an indication of possible grouping structures in the observed features themselves. The proposed model is then defined as follows 
\begin{eqnarray}\label{prop_model}
x_i &=& Z\Lambda_c u_i + \varepsilon_i \nonumber \\
&=& \tilde{\Lambda}u_i + \varepsilon_i \hspace*{1cm} i = 1, \dots, n,
\end{eqnarray} 
with $x_i, u_i$ and $\varepsilon_i$ previously defined while $Z = \{z_j\}_{j=1,\dots,p}$, with $z_j = (z_{j1}, \dots, z_{jG})$, is a $p \times G$ latent allocation matrix where $G$ is the number of variable clusters. Here the standard binary partition is adopted for $Z$; therefore $z_{jg} = 1$ if the $j$-th variable belongs to the $g$-th group and 0 otherwise. Lastly, $\Lambda_c = \{ \Lambda_{c,g}\}_{g=1,\dots,G}$, with $\Lambda_{c,g} = (\lambda_{c,g1},\dots,\lambda_{c,gK})$, is a $G \times K$ matrix whose $g$-th row contains the unique and representative loading values for the $g$-th variable cluster. Note that, as a direct consequence of the specification of the model (\ref{prop_model}), $\tilde{\Lambda}$ has duplicate row values. We believe this constitutes a sensible way to account for redundancy in the observed features, by simply constraining the relations with the latent factors to be equal for those variables belonging to the same cluster. 

The distributional properties highlighted above for the standard FA model are still valid.  Therefore, we have that $(x_i | u_i, z) \sim \mathcal{N}_p(\tilde{\Lambda}u_i, \Psi)$ while $(x_i|z) \sim \mathcal{N}_p(0, \tilde{\Sigma})$ where $\tilde{\Sigma} = \tilde{\Lambda}\tilde{\Lambda}^T + \Psi$. It is straightforward to see how the proposed model induces an even more parsimonious decomposition of the covariance matrix. In fact the specification of our model entails a possibly drastic reduction in the total number of covariance parameters to estimate; 
for model (\ref{prop_model}) this number is equal to $(G \times k) + p$, for model (\ref{eq:stdFA}) it is equal to $(p \times k) + p$; due to rotational invariance in FA, the number of identifiable parameters is fewer than this. Clearly, the smaller the number of variable clusters $G$, hence the more redundancy is observed in the data, the greater will be the reduction. 

From an interpretative point of view the estimation of the allocation matrix $Z$ allows obtaining a clustering of the variables. This partition gives insights into the redundancy phenomenon under study by clearly highlighting which variables are strongly correlated and providing similar information. 

Finally, note that our proposal may be adapted in order to detect both redundant and uninformative variables by a priori forcing all the elements in a single specific row of $\Lambda_c$ to be exactly equal to zero. This would imply that all the variables assigned to the corresponding group are modelled as noise and are uncorrelated with all the other variables. 

\subsection{Likelihood and prior specification}\label{sec:likelihood_prior}
Under the specification of model (\ref{prop_model}), and recalling that $(x_i | u_i, z) \sim \mathcal{N}_p(Z\Lambda_c u_i, \Psi)$, the corresponding likelihood function is given by 
\begin{eqnarray}\label{eq:likelihood}
\mathcal{L}(X | \Lambda_c,\Psi,Z, U) &=& \prod_{i = 1}^n (2\pi)^{-\frac{p}{2}}|\Psi|^{-\frac{1}{2}}\exp\left\{-\frac{1}{2} (x_i - Z\Lambda_c u_i)^T \Psi^{-1}(x_i - Z\Lambda_c u_i) \right\} \nonumber \\
&\propto& |\Psi|^{-\frac{n}{2}} \exp\left\{ -\frac{1}{2}\text{tr}\left[ \Psi^{-1}(X-U\Lambda_c^TZ^T)^T(X-U\Lambda_c^TZ^T) \right] \right\}
\end{eqnarray}
where $U = \{u_i\}_{i=1,\dots,n}$, with $u_i = (u_{i1},\dots,u_{iK})$, is the $n \times K$ factor scores matrix while $X, \Lambda_c, Z$ and $\Psi$ are defined as in the previous section. Lastly $|A|$ and $\text{tr}[A]$ denotes respectively the determinant and the trace of a generic matrix $A$.

Different strategies might be adopted to estimate the parameters involved in (\ref{prop_model}). From a frequentist perspective, the maximum likelihood estimates are usually obtained via iterative algorithms such as the ones proposed in \citet{joreskog1967some}, \citet{jennrich1969newton} and \citet{rubin1982algorithms}. Conversely, in this work we adopt a Bayesian approach to factor analysis estimation \citep[see e.g.][]{press1989bayesian,arminger1998bayesian,song2001bayesian}. More specifically, we assume independent prior distributions for the model parameters as follows 
\begin{eqnarray}\label{eq:priordist}
\Lambda_{c,g} &\sim& \mathcal{N}_K(0, \sigma^2_\lambda \mathbb{I}_K) \hspace{1.7cm} \text{for} \;\; g = 1, \dots,G \\
u_i &\sim& \mathcal{N}_K(0, \mathbb{I}_K) \hspace{2.12cm} \text{for} \;\; i = 1, \dots,n \\
\psi_j &\sim& \text{IG}(\alpha, \beta_j) \hspace{2.32cm} \text{for} \;\; j = 1,\dots, p \\
z_j &\sim& \text{PPM}(\alpha_z) \hspace{2.32cm} \text{for} \;\; j = 1, \dots, p \label{prior:ppm}.
\end{eqnarray}
The choice of the hyperparameters for the inverse gamma prior on the uniquenesses is guided by the suggestions in \citet{fruhwirth2010parsimonious}; here the authors avoid encurring in the Heywood problem by choosing $\alpha$ and $\beta_j$ so that $\psi_j$ tends to be bounded away from zero. More specifically in our analyses we set $\alpha = 2.5$ and $\beta_j = (\alpha - 1)/S^{-1}_{jj}$ where $S^{-1}$ represents the inverse of the sample covariance matrix. On the other hand $\sigma^2_{\lambda}$ might be chosen to be subjectively large in order to consider an uninformative prior for the rows of $\Lambda_c$. 

Some words of caution are required for the prior in (\ref{prior:ppm}). Let ${\bf c}$ be a clustering of indices $\{1,\dots, p\}$; even if different representations might be possible, we consider ${\bf c} = \{C_1,\dots,C_G \}$ as a collection of disjoint subsets such that $C_g$ contain all the indices of the variables belonging to cluster $g$-th. A product partition model \citep[PPM, ][]{hartigan1990partition,barry1992product} assumes that the prior probability for ${\bf c}$ is expressed as follows 
\begin{eqnarray}\label{eq:ppm}
\pi({\bf c} = \{C_1,\dots,C_G \}) \propto \prod_{g=1}^G \rho(C_g)
\end{eqnarray}
where $\rho(\cdot)$ is known as the \emph{cohesion function}. Since we have a one-to-one correspondence between the representation of the partition ${\bf c}$ as a collection of blocks and the one via the allocation matrix $Z$, with a slight abuse of notation, we specify the prior as in (\ref{prior:ppm}) even in our framework. Several different specifications for $\rho(\cdot)$ have been proposed in literature: in this work we consider $\pi({\bf c}) \propto \alpha_z^G \prod_{g=1}^G (|C_g| - 1)!$, where $|C_g|$ denotes the cardinality of the $g$-th cluster, sharing strong connections with the Dirichlet process that is widely used in the Bayesian clustering framework \citep{quintana2003bayesian}. Note that in our analyses we set $\alpha_z =1$ as is standard; however, this choice did not seem influential. When specifying a prior over the set of the partitions, another reasonable approach to take within our framework would consist of borrowing ideas from the Bayesian spatial clustering literature or to consider some additional information such as distances between the objects to be clustered \citep[see e.g.][and references therein]{blei2011distance,page2016spatial,dahl2017random,wehrhahn2020bayesian}. In such a way contiguous groups of wavelengths would be favored leading to some advantages in the modelling process for some specific applications. 

Given the likelihood function in (\ref{eq:likelihood}) and the specification of the priors outlined above, the posterior distribution is defined as follows
\begin{eqnarray}\label{eq:posterior}
\pi(\Lambda_c, \Psi, Z, U | X) &=& \mathcal{L}(X|\Lambda_c,\Psi, Z, U)\pi(\Lambda_c | \sigma^2_\lambda)\pi(\Psi | \alpha, \beta_j)\pi(Z | \alpha_z) \pi(U) \\
&\propto& \mathcal{L}(X|\Lambda_c,\Psi, Z, U)\prod_{g=1}^G \phi^{(K)}(\Lambda_{c,g}; 0, \sigma^2_\lambda\mathbb{I}_K)\prod_{j=1}^p \text{IG}(\alpha, \beta_j) \nonumber \\
&& \times \, \alpha_z^G \prod_{g=1}^G (|C_g| - 1)! \prod_{i=1}^n \phi^{(K)}(u_i; 0, \mathbb{I}_K)\nonumber 
\end{eqnarray}
where $\phi^{(K)}(\cdot, \mu, \Sigma)$ denotes the pdf of a $K$-dimensional Gaussian random variable with mean vector $\mu$ and covariance matrix $\Sigma$.

\subsection{Model estimation}\label{sec:modest}
Model parameter estimation is carried out in a Bayesian framework, using Markov Chain Monte-Carlo. Due to the conditionally conjugate nature of various prior distributions adopted, samples from (\ref{eq:posterior}) are obtained using a Gibbs sampling scheme with the exception of the latent variable allocation matrix $Z$ which is sampled via a Metropolis-Hastings step. The full conditional distributions are listed below (more details on their derivation and on the involved parameters are provided in the \hyperref[appendix]{Appendix}).
\begin{eqnarray}
 \text{vec}(\Lambda_c)|\dots &\sim& \mathcal{N}_{G\times K}(\mu_\lambda, \Sigma_\lambda) \label{eq:fullcond_lambda}  \\
 u_i|\dots &\sim& \mathcal{N}_K(\mu_u, \Sigma_u) \label{eq:fullcond_scores} \\
 \psi_j|\dots &\sim& \text{IG}(\alpha + n/2, \beta_j^*) \label{eq:fullcond_uniqueness}
\end{eqnarray}  
For the Metropolis-Hastings step to sample the  allocation matrix $Z$, we adapt to our case one of the moves proposed by \citet{nobile2007bayesian} in the so called \emph{allocation sampler}. Each single move attempts to reallocate to cluster $g_2$ a group of variables previously assigned to cluster $g_1$; in this way, by possibly reallocating blocks of variables, big moves are proposed so that the space will be explored faster. The detailed steps of the procedure are outlined hereafter:
\begin{enumerate}
  \item Draw, from the total $G$ variable clusters, a group $g_1$. If $n_{g_1} = |C_{g_1}| = 0$, where $|C_{g_1}|$ denotes the cardinality of group $g_1$, the move fails; 
  \item Compute $d(\Lambda_{c,g_1},\Lambda_{c,g'})$, the Euclidean distance between $\Lambda_{c,g_1}$ and $\Lambda_{c,g'}$, $\forall \; g' \in \{1,\dots,G\}$ with $g' \ne g_1$. Afterwards a second group $g_2$ is drawn from the set $\{1, \dots, G\} \setminus g_1$ with $\mathbb{P}(g' = g_2) \propto d(\Lambda_{c,g_1},\Lambda_{c,g'})^{-1}$; 
  \item Draw $M$ from the set $\{1, \dots,  n_{g_1}\}$ with $\mathbb{P}(M=m) \propto 1/m, \; \forall m = 1, \dots, n_{g_1}$;
  \item Select randomly $M$ observations among the $n_{g_1}$ belonging to group $g_1$ and reallocate them to group $g_2$;
  \item Denote with $Z$ and $Z'$ respectively the starting latent allocation matrix and the one after the reallocation move. The move itself is then accepted with probability $\min\{1,R\}$, where R is given by 
  \begin{eqnarray*}
  R = \frac{\pi(\Lambda_c, \Psi, Z', U | X) }{\pi(\Lambda_c, \Psi, Z, U | X) }\frac{\mathbb{P}(Z' \rightarrow Z)}{\mathbb{P}(Z \rightarrow Z')} \; .
  \end{eqnarray*}
  It can be easily shown that the proposal ratio is 
  $$
  \frac{\mathbb{P}(Z' \rightarrow Z)}{\mathbb{P}(Z \rightarrow Z')} = \frac{\sum_{m=1}^{n_{g_1}} \frac{1}{m}}{\sum_{m=1}^{n_{g_2}+M} \frac{1}{m}}\frac{n_{g_1}!n_{g_2}!}{(n_{g_1} - M)!(n_{g_2} + M)!} \;\;\; .
  $$
\end{enumerate}
Our modification of the procedure proposed by \citet{nobile2007bayesian} consists in changing the probabilities involved in the selection of $g_2$ and $M$. The rationale lies in the need to propose, at each step, the reallocation of blocks of variables while increasing the acceptance ratio by proposing moves involving similar clusters and by keeping a reasonable size of the blocks. 

Lastly, note that the model we are proposing, having a factor analytic structure, inherits standard identifiability issues related to the rotational invariance property. A common solution consists in considering some constraints on the factor loadings \citep[see e.g.,][]{arminger1998bayesian,lopes2004bayesian}. In our framework, where the factor analytic structure of the model may be seen as a tool to reconstruct $\Sigma$ in a parsimonious way, identification is not strictly necessary. Moreover it has been noted \citep{bhattacharya2011sparse} that identifiability constraints may lead to order dependence among the variables and general inefficiencies. As a direct consequence we decided not to consider such constraints in our modelling strategy. 

\subsection{Model selection}\label{sec:modelselection}
In the previous sections, the number of factors $K$ has been considered as fixed; in practice inference on $K$ constitutes one of the most challenging and trickiest issues to tackle when considering factor analytic models. Several different approaches have been proposed in literature, with a standard one resorting to widely known information criteria, such as the AIC and BIC, as selection tools. Nonetheless these criteria might not be reliable in high-dimensional settings and they are not theoretically justified in the FA framework. From a Bayesian perspective it is worth mentioning the work by \citet{lopes2004bayesian} where the authors propose a reversible jump Markov Chain Monte-Carlo algorithm, moving between models having different number of factors. Another viable approach comes from the nonparametric literature where models with an infinite number of factors have been widely used in combination with shrinkage priors on the loading matrices \citep{bhattacharya2011sparse,durante2017note,schiavon2020truncation}; such a strategy allows the automatic choice of the number of the active factors, being the ones with non-negligible loading values. 

Note that in the framework developed herein, the model selection step is even more troublesome since the choice of $K$ is coupled with the one of the number of variable clusters $G$. A similar problem arises in the context of mixture of factor analyzers \citep{ghahramani1996algorithm} where usually different models, corresponding to different configurations of factors and mixture components, are compared by means of information criteria. In a Bayesian framework a solution has been proposed by \citet{fokoue2003mixtures} where the authors adopt a stochastic model search to jointly select the optimal number of clusters and factors. 

In order to address the issues mentioned above, in the considered settings different strategies may be adopted. A viable alternative to the information criteria usually adopted and more coherent with the estimation routine outlined in Section \ref{sec:modest}, may consist of considering the BICM (BIC Monte-Carlo) or the AICM (AIC Monte-Carlo) proposed by \citet{raftery2006estimating} and defined as $\text{BICM} = 2\log\tilde{\mathcal{L}} - 2s^2_l \log(n)$ and $\text{AICM} = 2\log\overline{\mathcal{L}} - 2s^2_l$ where $\tilde{\mathcal{L}}$, $\overline{\mathcal{L}}$ and $s^2_l$ are respectively the maximum observed, the mean and the variance of the likelihood computed for each posterior samples. Another, somewhat heuristic, approach is given by the so called BIC-MCMC \citep{fruhwirth2011label} with $\text{BIC-MCMC} = 2\log\tilde{\mathcal{L}} - \nu \log(n)$ where $\nu$ is the total number of parameters in the model. Note that, not depending on $s^2_l$, the BIC-MCMC turns out to be less influenced by possible fluctuations and jumps in the log-likelihood values across the MCMC draws. 

Furthermore, an exhaustive search of the model space is computationally expensive, if not infeasible, considering that in our scenario both $K$ and $G$ might span over a wide range of values. Moreover, in the given framework, the focus is on models providing good and parsimonious reconstructions of the covariance matrices, jointly with indications about which variables provide similar information through redundancy, rather than on finding the optimal number of factors and groups. For these reasons, in this work, we consider an ad hoc initialization strategy which yields a promising configuration $(K_{\text{init}}, G_{\text{init}})$ for the number of factors and variable clusters. The procedure consists in the following steps: 
\begin{enumerate}
  \item Estimate a standard FA model as defined in (\ref{eq:stdFA}) for $k = 1, \dots, K_{\text{max}}$, with $K_{\text{max}}$ chosen sufficiently large. This yields the loading matrices $\Lambda_k$, $k=1, \dots, K_{\text{max}}$;
  \item Use a model-based clustering strategy (see \citet{fraley2002model} or \citet{bouveyron2019model} for a recent review) to obtain a partition of the rows of $\Lambda_k$, for $k = 1, \dots, K_{\text{max}}$, into $G_k$ groups with $G_k$ automatically selected by means of the Bayesian Information Criterion (BIC);
  \item Build new loading matrices $\overline{\Lambda}_k$, for $k = 1, \dots, K_{\text{max}}$, where the rows of $\Lambda_k$ are replaced with the mean of the cluster they belong to. In such a way the repeated row values structure of $\tilde\Lambda$ in (\ref{prop_model}) is mimicked; 
  \item Considering the distributional properties of FA models, compute the BIC for all the models corresponding to different configurations $(k,G_k)$, with $k = 1, \dots, K_{\text{max}}$. Select as $(K_{\text{init}}, G_{\text{init}})$ the configuration which attains the highest value for the BIC. 
\end{enumerate}
This approach allows to find reasonable values that might be used as the starting point of a local search. 
More specifically, once $(K_{\text{init}}, G_{\text{init}})$ are obtained by running the initialization procedure illustrated above, we consider a greedy search model selection strategy. From a practical point of view we fit four different models corresponding to $(K_{\text{init}} \pm 1, G_{\text{init}} \pm 1)$ and we compare them by means of the BIC-MCMC. The model with the best value of the information criterion is then selected and its neighboring models are subsequently estimated. These two steps are iterated until no improvements in the BIC-MCMC values are found. The best model according to the BIC-MCMC is then selected and subsequently used to reconstruct the covariance structure and to obtain a partition of the wavelengths.

Lastly, note that some sensitivity analyses conducted and reported in the next section confirms that running a global and exhaustive search is not strictly necessary if covariance reconstruction is the final aim. 

\section{Synthetic data}\label{sec:simulations}
In this section we investigate the performances of the proposed procedure on some synthetic datasets. The aim of the analyses reported hereafter is twofold. On one hand we want to quantify the deterioration of the results when a wrong model, having different $(K,G)$ values with respect to the model generating the data, is employed. The possible deterioration is studied in terms of variable partitions quality, that is measured according the \emph{Adjusted Rand Index} \citep[ARI,][]{hubert1985comparing}, and of correlation reconstruction. The latter is evaluated considering two different criteria measuring the dissimilarity between the true correlation matrix $R = D^{-1/2}\Sigma D^{-1/2}$ and the estimated one $\hat{R} = \hat{D}^{-1/2}\hat{\Sigma}\hat{D}^{-1/2}$ with $D = \text{diag}(\sigma^2_1,\dots,\sigma^2_p)$ and $\hat{D} = \text{diag}(\hat{\sigma}^2_1, \dots, \hat{\sigma}^2_p)$. The first criterion we considered is the \emph{Mean Squared Error} (MSE) that is defined as
\begin{eqnarray*}
\text{MSE}(R,\hat{R}) = \frac{1}{\frac{p(p+1)}{2}} \sum_{j=1}^p \sum_{j \ge j'} (R_{jj'} - \hat{R}_{jj'})^2
\end{eqnarray*}
where $R_{jj'}$ represents the $(j,j')$-th element of the matrix $R$. The second criterion adopted is the RV coefficient \citep{abdi2007rv} expressed as  
\begin{eqnarray*}
\text{RV}(R,\hat{R}) = \frac{\text{tr}(R^T \hat{R})}{\sqrt{\text{tr}(R^T R)\text{tr}(\hat{R}^T\hat{R})}}
\end{eqnarray*}
and taking values between 0 and 1 where values closer to 1 denote a greater similarity between the matrices. Note that we consider the correlation matrices and not the covariance ones in order to have more interpretable indications from a MSE perspective. \\ 
On the other hand the second aim of the simulation study consists in the numerical exploration of the quality of the initialization strategy proposed in Section \ref{sec:modelselection} in order to find promising configurations for the number of factors $K$ and variable clusters $G$.

A total of $B = 200$ samples have been drawn with sample size $n = 500$ and $p=40$ variables. The data are generated according to the probabilistic mechanism underlying model (\ref{prop_model}) so that the sampled vectors $x_i$'s are distributed as a Gaussian random variables with zero mean and covariance matrix $\Sigma_{\text{true}} = \Lambda_{\text{true}}\Lambda_{\text{true}}^T + \Psi_{\text{true}}$. The true number of factors and of variable clusters have been fixed to $K_{\text{true}} = 3$ and $G_{\text{true}} = 5$ respectively. Prior to running the Gibbs sampler outlined in Section \ref{sec:modest}, the involved parameters are initialized by estimating a standard FA model where the factor loadings are obtained as the cluster centroids of a $k$-means clustering procedure, which also allows to obtain the starting values for the loadings partition. The hyperparameters have been selected according to the indication given in Section \ref{sec:likelihood_prior} with $\sigma_\lambda = 5$ to entail a priori uninformativeness about the dispersion of the factor loadings. All the reported analyses have been conducted within the \texttt{R} environment \citep{R-software} with the aid of the \texttt{mclust} package \citep{scrucca2016mclust}.

\begin{table}[!t]
\centering 
\caption{Means, over the $B$ simulated samples, of the ARI values (and their standard errors) comparing the true and the estimated covariance matrices for varying values of $K$ and $G$. Bold cell represents the true model generating the data.}
 \begin{tabular}{cccccc}
  \hline
 $K$ | $G$ & 3 & 4 & 5 & 6 & 7 \\ 
  \hline
2& 0.728 (0.140) & 0.915 (0.082) & 0.985 (0.044) & 0.993 (0.029) & 0.995 (0.024) \\ 
3& 0.709 (0.155) & 0.890 (0.109) & {\bf 0.970 (0.063)} & 0.982 (0.054) & 0.991 (0.034) \\ 
4& 0.719 (0.140) & 0.861 (0.121) & 0.934 (0.101) & 0.962 (0.058) & 0.970 (0.051) \\ 
5& 0.696 (0.162) & 0.837 (0.141) & 0.909 (0.097) & 0.918 (0.093) & 0.936 (0.076) \\ 
   \hline
\end{tabular}
\label{tab:tab_ari}
\end{table}

Results are reported in Tables \ref{tab:tab_ari}, \ref{tab:tab_mse}, \ref{tab:tab_rv} and \ref{tab:tab_selection}. First of all note that variable partitions might be erroneously seen as a byproduct of the procedure proposed in Section \ref{sec:model_definition}, helping to reduce even further the number of free parameters when resorting to factor analysis. Nonetheless obtaining variable clusters can represent the final aim of the analyses since it produces relevant insights about the phenomenon that we are studying, as it will be clear for the application in Section \ref{sec:application}. For this reason a proper evaluation of the clustering performances, even in simulated scenarios, is crucial to validate our procedure.
From Table \ref{tab:tab_ari} we can see how in these scenarios the obtained variable partitions are generally close to the true groupings, as the ARI generally achieves good values. A more careful investigation of the results shows how, as expected, the values are strongly influenced by the number of cluster $G$ used in the model fitting procedure. Nonetheless this behavior turns out not being symmetrical since, while an underestimation of the number of clusters appears to be harmful, overestimating $G$ looks harmless if not beneficial; this might give a clue about the spuriousness of some of the clusters when $G > G_{\text{true}}$. The impact of the number of factors is less evident as the ARI values appear robust with respect to changes in $K$. A relevant behavior to highlight consists in the slight degradation of the results when increasing the number of factors; note that, as $K$ increases, the dimensionality of the space in which cluster searches are conducted increases too possibly enhancing the sparsity of the data and deteriorating clustering performances.  

\begin{table}[t]
\centering
\caption{Mean, over the $B$ simulated samples, of the MSE (and their standard errors) comparing the true and the estimated correlation matrices for varying values of $K$ and $G$. Bold cell represents the true model generating the data.}
\begin{tabular}{cccccc}
  \hline
 $K$ | $G$ & 3 & 4 & 5 & 6 & 7 \\ 
  \hline
2&  0.052 (0.048) & 0.022 (0.030) & 0.009 (0.010) & 0.009 (0.010) & 0.010 (0.011) \\ 
3& 0.064 (0.060) & 0.019 (0.030) & {\bf 0.004 (0.011)} & 0.002 (0.010) & 0.002 (0.007)\\ 
4& 0.062 (0.059) & 0.027 (0.031) & 0.014 (0.030) & 0.007 (0.015) & 0.004 (0.010) \\ 
5& 0.066 (0.064) & 0.032 (0.041) & 0.015 (0.024) & 0.013 (0.026) & 0.008 (0.020) \\ 
   \hline
\end{tabular}
\label{tab:tab_mse}
\end{table}

\begin{table}[!t]
\centering 
\caption{Means, over the $B$ simulated samples, of the RV coefficients (and their standard errors) comparing the true and the estimated correlation matrices for varying values of $K$ and $G$. Bold cell represents the true model generating the data.}
 \begin{tabular}{cccccc}
  \hline
 $K$ | $G$ & 3 & 4 & 5 & 6 & 7 \\ 
  \hline
2& 0.911 (0.089) & 0.972 (0.031) & 0.989 (0.019) & 0.987 (0.021) & 0.984 (0.025) \\ 
3& 0.895 (0.097) & 0.968 (0.052) & {\bf 0.993 (0.020)} & 0.997 (0.006) & 0.997 (0.011) \\ 
4& 0.898 (0.095) & 0.960 (0.049) & 0.982 (0.035) & 0.992 (0.015) & 0.994 (0.009) \\ 
5& 0.894 (0.104) & 0.953 (0.057) & 0.979 (0.035) & 0.985 (0.029) & 0.989 (0.022) \\ 
   \hline
\end{tabular}
\label{tab:tab_rv}
\end{table}

In Tables \ref{tab:tab_mse} and \ref{tab:tab_rv} the results concerning correlation matrix reconstruction are reported. First of all it is relevant to highlight how both the MSE and the RV coefficient tend to provide very similar indications. Generally speaking the obtained performances are good overall, regardless of the values of $K$ and $G$. A more detailed look reveals how, coherently with what happens for the clustering quality, underestimation of the number of variable clusters $G$ seeems to have a more visible impact on the degradation of the results. On the other hand overestimation of $G$ seems to have little impact on the values of the MSE and RV coefficient and the same holds for different choices of the number of factors.

Finally in Table \ref{tab:tab_selection} the performances of the model selection initialization strategy outlined in Section \ref{sec:modelselection} are displayed. A first promising result is given by the fact that the true model generating the data is the one selected more often by this strategy. Moreover the right number of factors is chosen in more than 80\% of the cases. On the other hand the correct number of clusters looks harder to detect, with $G= 5$ being selected one every two samples. Nonetheless a tendency to overestimate $G$ is witnessed with this behavior being generally beneficial according to the results commented above on the quality of the partitions and the correlation reconstruction.    

\begin{table}[!t]
\centering
\caption{Proportion of times, over $B=200$ simulated samples, when a specific configuration $(K,G)$ has been selected by the initialization strategy outlined in Section \ref{sec:modelselection}. Bold cell represents the true model generating the data.}
\begin{tabular}{cccccc}
  \hline
 $K$ | $G$ & 4 & 5 & 6 & 7 & 8 or more \\ 
  \hline
1& 0.005 & 0.000 & 0.000 & 0.000 & 0.000 \\
2& 0.030 & 0.010 & 0.005 & 0.000 & 0.005 \\ 
3& 0.125 & {\bf 0.460} & 0.115 & 0.040 & 0.080 \\ 
4& 0.000 & 0.005 & 0.010 & 0.000 & 0.035 \\ 
5& 0.000 & 0.000 & 0.005 & 0.000 & 0.005 \\ 
6 or more & 0.000 & 0.030 & 0.005 & 0.000 & 0.030 \\ 
   \hline
\end{tabular}
\label{tab:tab_selection}
\end{table}

We strongly believe that the results reported in this section have to be considered as a whole, in order to obtain useful indications about reasonable paths to take when analyzing real datasets. From Table \ref{tab:tab_selection} we can see how the proposed initialization strategy never selected $G < 4$ and how, when not selecting the true model generating the data, it favors the overestimation of both $K$ and $G$. These indications, if coupled with the ones obtained from Tables \ref{tab:tab_ari}, \ref{tab:tab_mse} and \ref{tab:tab_rv}, suggest that the initialization strategy tends to select models producing satisfactory performances both from a clustering and from a correlation reconstruction perspectives. As commented above a more pronounced deterioration of the results is witnessed especially when $G=3$ while, overestimation of both the number of clusters and factors, generally lead to an amelioration of the performances. As a consequence we believe that the proposed strategy might be fruitfully used as a fast and effective replacement of more intensive and time consuming grid searches over $K$ and $G$ coupled with the reliance to some information criterion that has to be carefully selected.  

In fact, note that some further analyses, not reported in the paper, generally showed the unreliability of the information criteria introduced in Section \ref{sec:modelselection}. As a general indication it has been noted that BIC-MCMC is more stable with respect to AICM and BICM since the latter ones could be strongly influenced by jumps, usually due to updates of the allocation matrix $Z$, of the likelihood at a given step of the Gibbs updating procedure. Nonetheless in our analyses all the three criteria selected the true model generating the data less often with respect to the suggested initialization strategy.


\section{Application to the milk MIR spectroscopy data}\label{sec:application}
In this section, the proposed method is applied to the milk MIR spectroscopy data described in Section  \ref{sec:data_description}. The initialization of the parameters involved in the model and the specification of the hyperparameters have been carried out coherently with what we have done for the synthetic data in the previous section. Moreover, prior to running the proposed metholodogy, we removed from each single spectrum three wavelength regions supposed to be highly noisy \citep{hewavitharana1997fourier} namely the ones from  1592 cm$^{-1}$ to 1720 cm$^{-1}$, from 2996 cm$^{-1}$ to 3698 cm$^{-1}$ and from 3818 cm$^{-1}$ to 5010 cm$^{-1}$. Consequently we end up working with a dataset having $n= 4320$ milk samples and $p = 533$ wavelengths measured.

The initialization procedure outlined in Section \ref{sec:modelselection} selects different $(K,G)$ configurations for the Pasture and for the TMR samples. More specifically, in the first case it selects a number of factors $K$ equal to 4 and a number of variable clusters $G$ equal to 25 while in the latter one $K = 3$ and $G = 19$. This might give a first rough indication about the more complex wavelengths relations underneath the samples coming from pasture fed cows since in order to capture those relations an higher number of clusters, lying in an higher dimensional reduced subspace, is needed.

\begin{figure}[!t]
\centering
\begin{minipage}{.47\textwidth}
  \centering
  \includegraphics[height = 6cm, width=6cm]{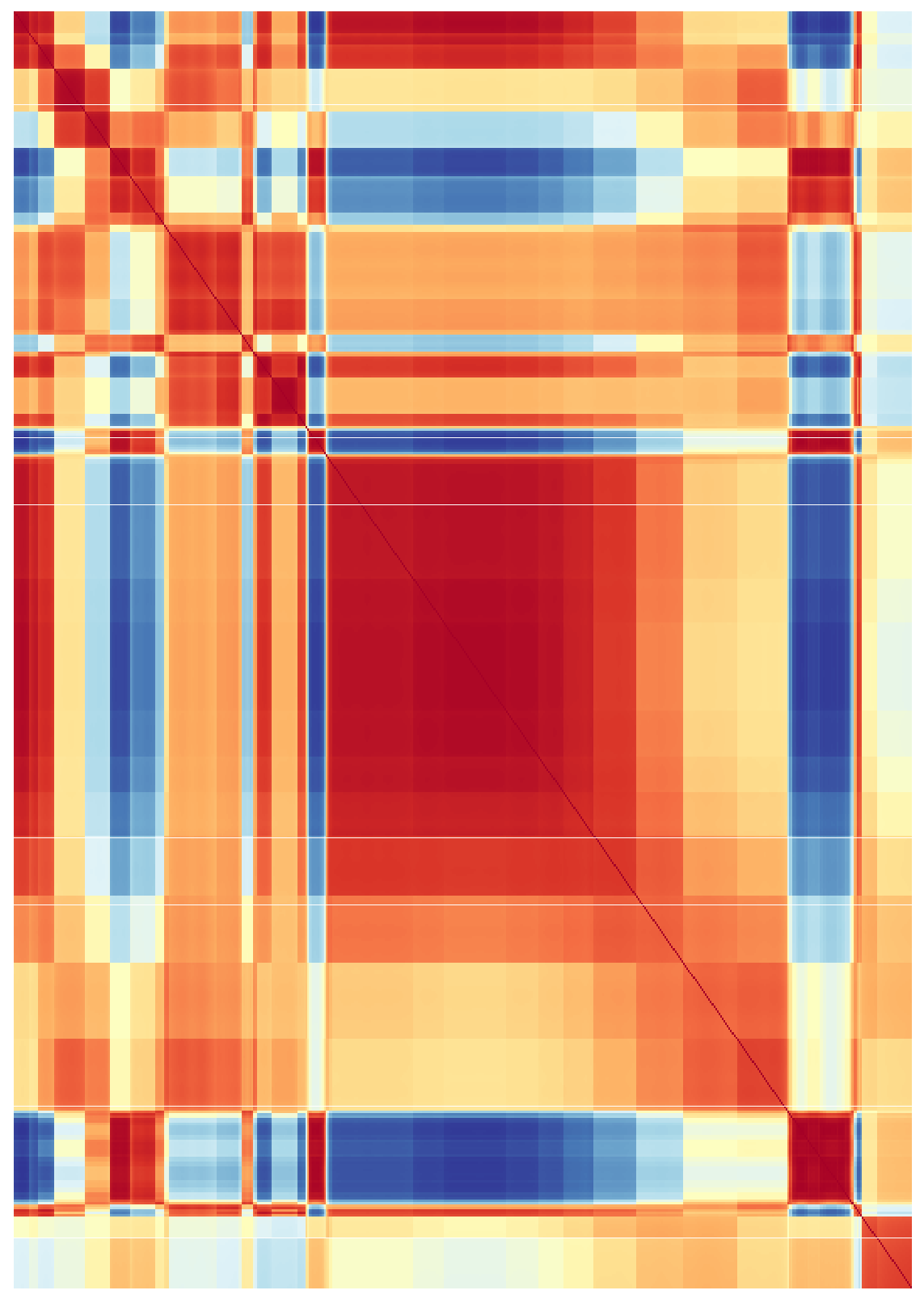}
\end{minipage}%
\begin{minipage}{.47\textwidth}
  \centering
  \includegraphics[height = 6cm,width = 6.5cm]{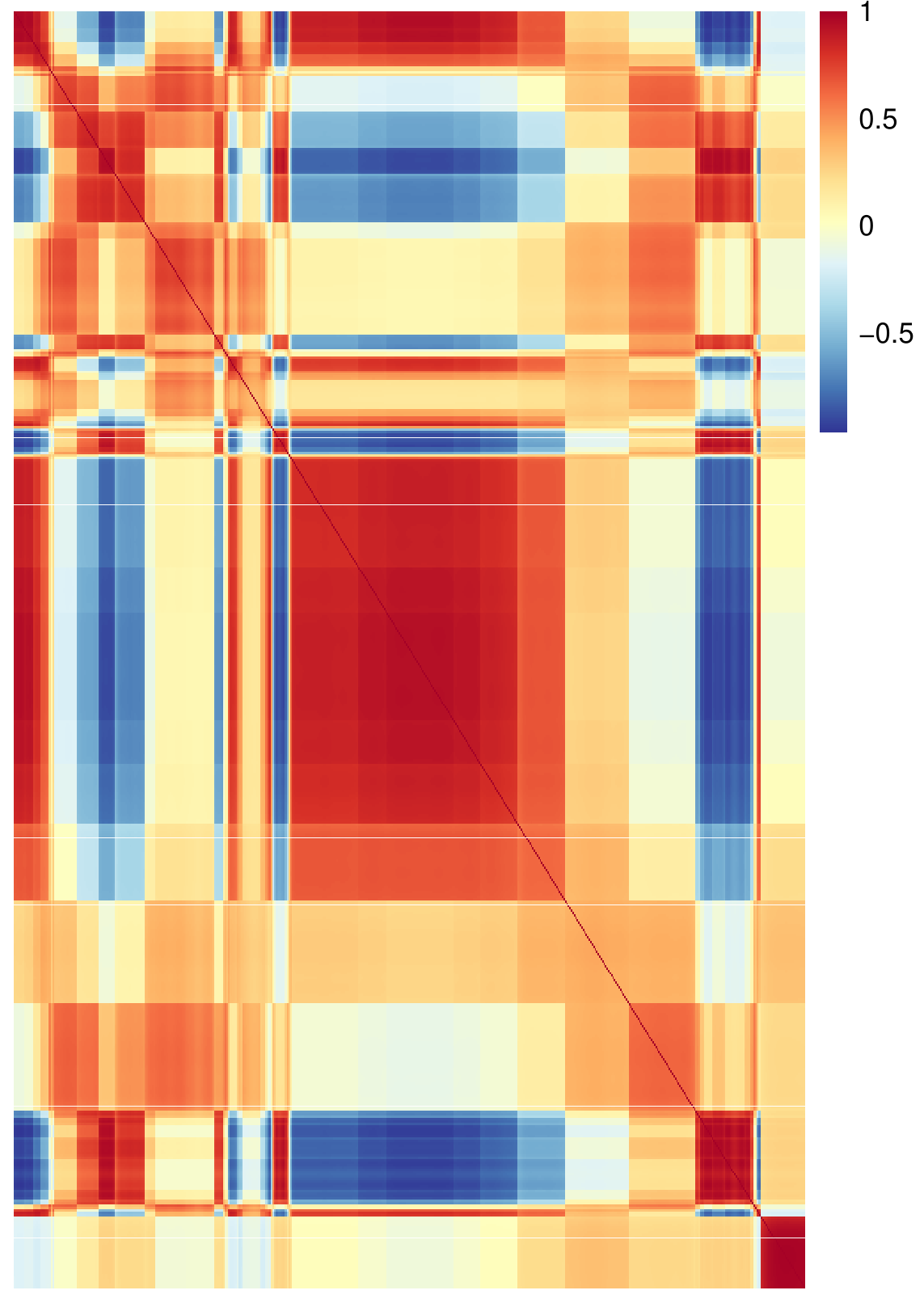}
\end{minipage}
  \caption{Estimated correlation matrices computed on the milk samples produced by pasture fed cows (on the left) and TMR fed cows (on the right).}
 \label{fig:estCor}
\end{figure}

In Figure \ref{fig:estCor} the estimated correlation matrices, obtained running the proposed model with the mentioned $(K,G)$ values, are reported. By comparing these matrices with the sample correlation ones displayed in Figure \ref{fig:sampleCor} we can obtain an indication about the capabilities of our methodology to map the data into lower dimensional subspaces while retaining the relevant correlation structures. Denoting with $R$ the sample correlation matrix and with $\tilde{R}$ the estimated one we have that, for the pasture samples, $\text{MSE}(R_{\text{Pasture}},\tilde{R}_{\text{Pasture}}) = 0.021$ and $\text{RV}(R_{\text{Pasture}},\tilde{R}_{\text{Pasture}}) = 0.980$. On the other hand, in the case of TMR, we obtain $\text{MSE}(R_{\text{TMR}},\tilde{R}_{\text{TMR}}) = 0.035 $ and $\text{RV}(R_{\text{TMR}},\tilde{R}_{\text{TMR}}) = 0.965$. These results suggest that our method reconstructs in quite a satisfactory way the relations among the wavelengths and that the initialization strategy selects reasonable values for $K$ and $G$. 

Furthermore, the graphical inspection of Figure \ref{fig:estCor}, suggests how our variable clustering mechanism tends to favour the appearance of blocky structures thus possibly simplifying the practical interpretation of the relations among wavelengths. Moreover, even in the face of a rather similar correlation structure, this characteristic of our proposal allows to highlight even more the differences in the correlation among different wavelengths regions between milk samples coming from pasture fed and TMR fed cows. These graphical differences may serve as an interesting starting point to study how the diet regimens can impact the chemical processes underlying the spectral behaviour. 

\begin{table}[!t]
\caption{Confusion matrix comparing wavelengths partitions obtained on samples from pasture fed (on the rows) and TMR fed (on the columns) cows. Blank spaces are used instead of zeroes.}
\centering
\scalebox{0.85}{
\begin{tabular}{cccccccccccccccccccc}
  \hline
 & 1 & 2 & 3 & 4 & 5 & 6 & 7 & 8 & 9 & 10 & 11 & 12 & 13 & 14 & 15 & 16 & 17 & 18 & 19 \\ 
  \hline
1 & 7 & 2 & & & & & &  &  & &  &  &  &  &  &  &  &  & \\ 
  2 &  &  &  15 &  &   &   1 &   2 &   &  &  &  &  &  &  & &  &  &  &  \\ 
  3 &  &   &  &  15 &  &  &  &  &  &  & &  &  &   &  &   &   &  &   \\ 
  4 &  &  &  &  &   6 &   &   1 &  5 &  & &  &  &  & &  &  &  &  &  \\ 
  5 &  &  &  &  &  &  29 &  &  &  &  &  &  &  &  &  &   3 &  &  &   \\ 
  6 &  &   &   &   &  &  &  &  &  20 &  &  &  &  &  &    &   &  &  &  \\ 
  7 &  &  &  &  &  &  15 &  &  &  &  &  &  &  &  &    &  &  &   &  \\ 
  8 &  &   & &  & &  &  12 &   &   &   3 &   &  &    &    &    &    &  &  &    \\ 
  9 &  &  & &  &  &   &    &    &  12 &    &    &    &    &  &  &   2 &   &  &   \\ 
  10 &   1 &   1 &  &  &  &  &  &  2 &  & 1 &  &  & &   &  &   &  &   & \\ 
  11 &  &  &  &  &   2 &  &  &   6 &  &   2 &  &  &  &  &  &  &  &  &  \\ 
  12 &  &   &  &  & &  &  &  &  &  &  55 &   8 &  &  & &  &  &  &\\ 
  13 &  &   4 &  &  &  &  &  &   1 &  &  &  &  &  37 &  &  &  &  &  & \\ 
  14 &  &  &  &  &  &  &  &  &   1 &  &  &  &  &  24 &  &  &  & &  \\ 
  15 &  & &  &  &  &  &  &  &  &  &  &  29 &   8 & &  & & &  & \\ 
  16 &   &  & &    &  &  &  &  &  &  &  15 &  &  &  &   5 &  &   &  &  \\ 
  17 &  &  & &  & &  &  & &  &  &  &  &  &  &  26 &  &  &  &  \\ 
  18 &  &  &  &  &  &  &  &  &  &   &   &  & &  &  & &   &  10 &  \\ 
  19 &  &  &  &  &  &  & &  &  &  &  &  &  &   &   3 &   &  27 & &  \\ 
  20 &   &    &   &   &   &   &  &  &   &   &   &   &   &   6 &   &  & &   3 &    \\ 
  21 &    &    &    &  &    &  &    &   &  &   &    &  &   &  &   &  33 &   &    &    \\ 
  22 &  &   &    &   &   &   &    &   &  &    &  &   &    &   &    &  15 &  17 &   &    \\ 
  23 &  &   &   &  &   &  &   &   &   &   &   &  &   &   &   &   &  &  11 &   \\ 
  24 &    &   &   &  &  &   &  &  &    &    &   &    &   &    &  &  & & &   9 \\ 
  25 &   &    &  &    &    &  &   & &  &  &  &  &  &  &   &    &    &   &  21 \\ 
   \hline
\end{tabular}}
\label{tab:table_clust_results}
\end{table}

\begin{figure}[!t]
\centering
\hspace*{-.5cm}
\includegraphics[height = 7cm, width = 13cm]{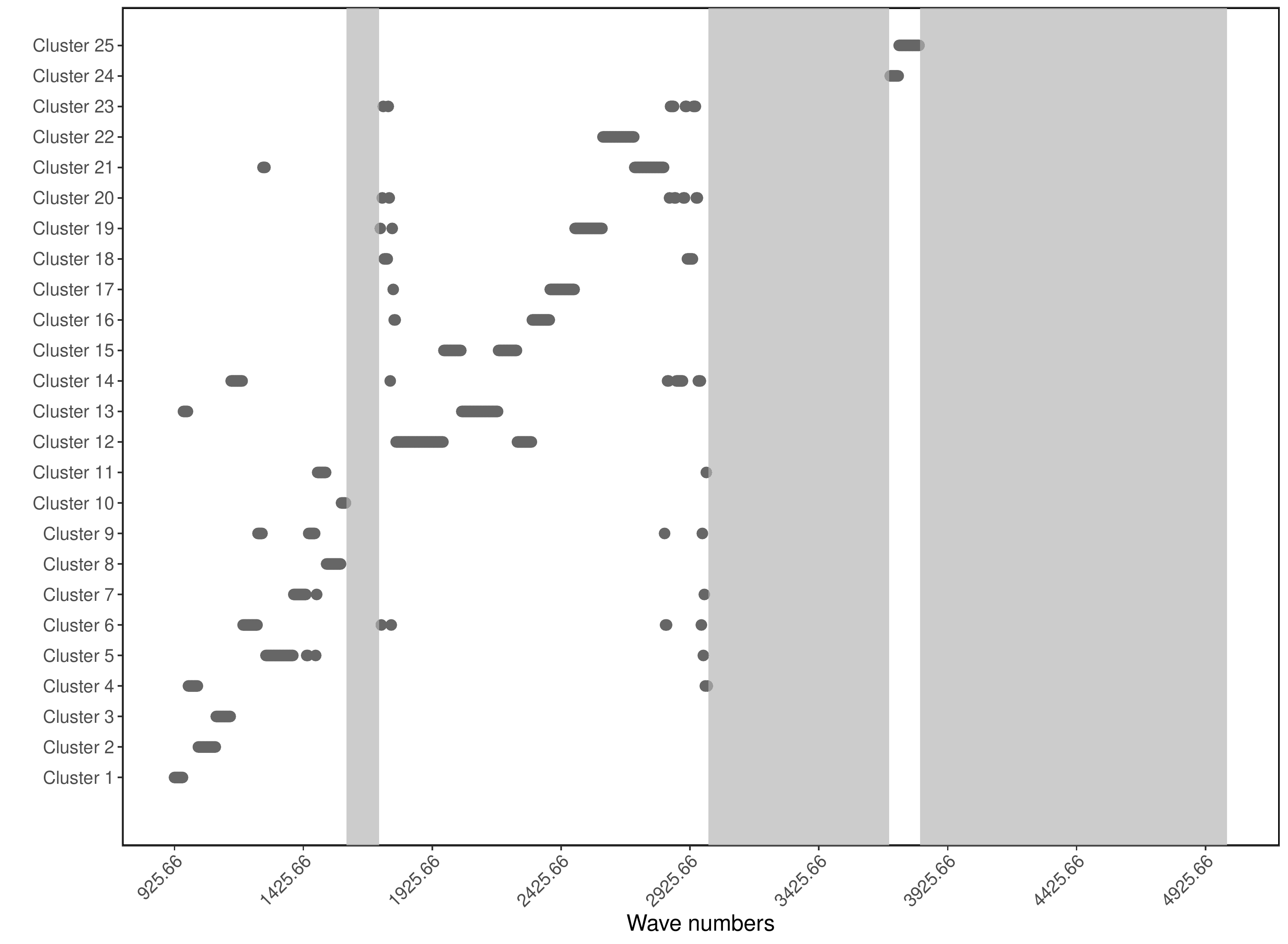} \\
\hspace*{-.5cm}
\includegraphics[height = 7cm, width = 13cm]{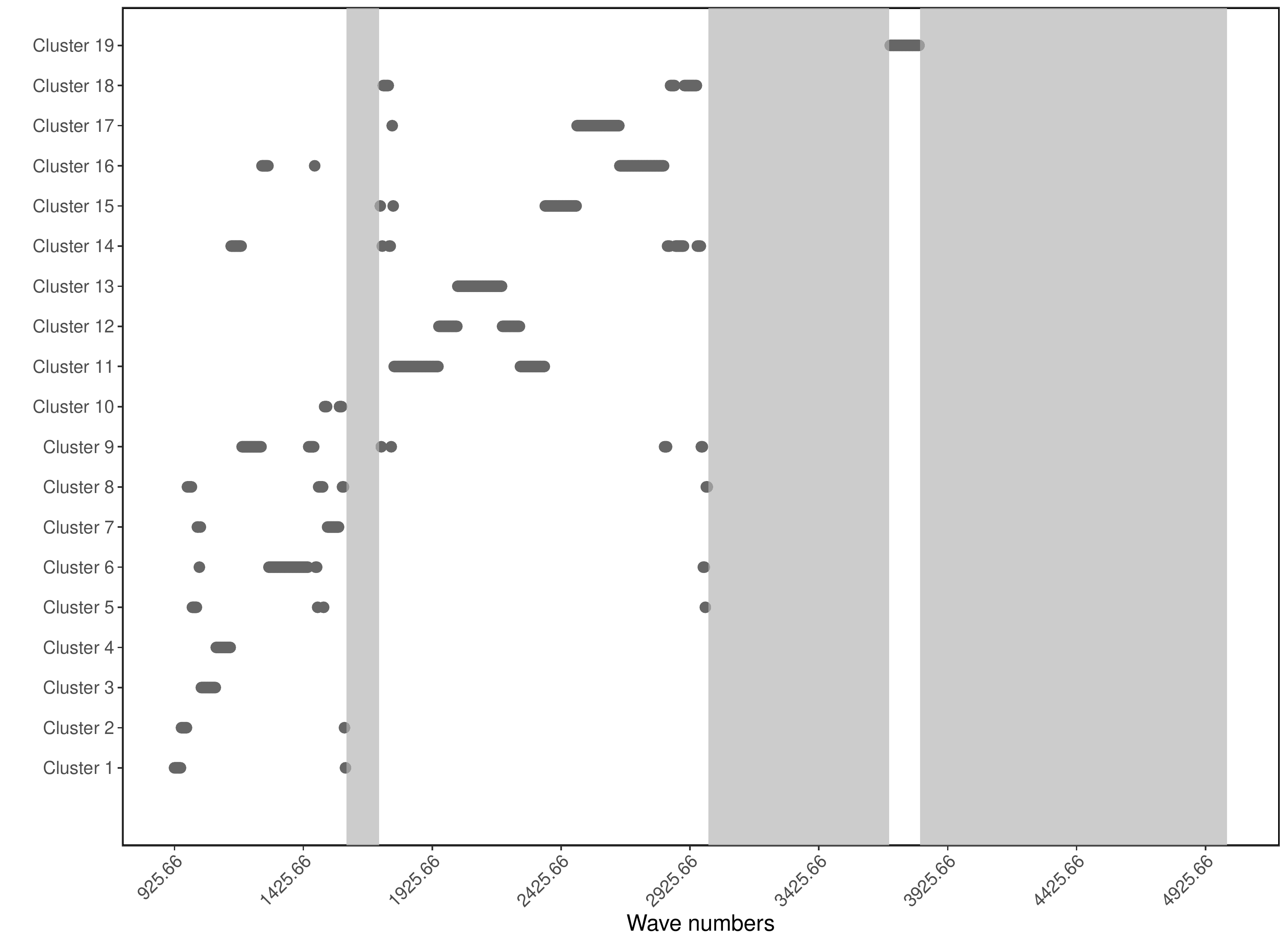}
\caption{Wavelengths partitions obtained on samples from pasture fed (on the top) and TMR fed (on the bottom) cows. The grey shaded areas correspond to the removed noisy regions.}
\label{fig:cluster_waves}
\end{figure}

Coherently with the comments we have made for the synthetic data analyses, another way we consider to assess the performances of our methodology consists in the investigation of the partitions of the variables. In Table \ref{tab:table_clust_results}, we report the confusion matrix comparing the two clusterings of the wavelengths obtained on the pasture and on the TMR milk samples. It stands out how, despite having a different number of clusters, the two partitions are similar as the table show an almost diagonal structure. A confirmation about their similarity is given by the quite high ARI value being equal to 0.651. The agreement between the two partitions is somehow expected since we are examining milk samples where the only different experimental condition consists in the different diet regimens. Moreover, this behaviour may be seen as a strong signal about the presence of a real clustering structure in the measured wavelengths, thus entailing a traceable redundancy in the information they provide. Nonetheless, a careful analysis of the results in Table \ref{tab:table_clust_results} reveals how the different number of clusters among the partitions generally imply that the large TMR variable clusters are split in two pasture variable clusters, as it happens for example for TMR clusters 11 and 17. This behaviour might provide some initial indications, possibly deserving further explorations, about how the different diets can impact chemical features in the milk in turn modifying the structures we see in the spectral data. Similar indications can be drawn from Figure \ref{fig:cluster_waves} where the partitions of the wavelengths for the two different diet regimens are visually represented. 

Furthermore, note that the insights obtained from a clustering perspective may be exploited to build variable selection tools possibly useful both for exploratory or graphical analyses and for classification purposes. In fact, the indication about the strong redundancy implied by the witnessed clustering structures can be used in order to build new features defined as summaries of the groups themselves possibly highlighting differences among pasture and TMR samples. 

\begin{figure}[!t]
\centering
\begin{minipage}{.55\textwidth}
  \centering
  \includegraphics[height = 8cm, width=6cm]{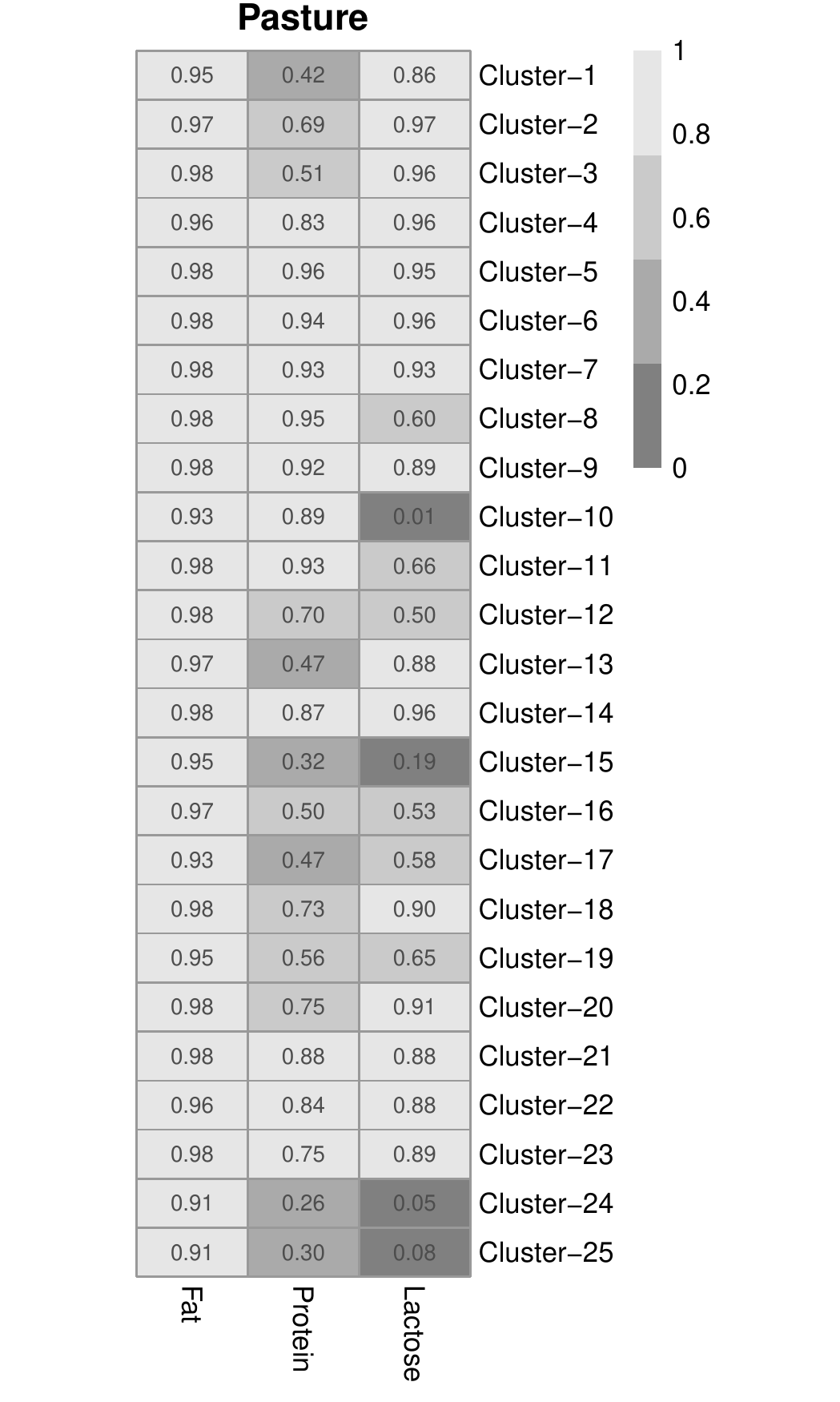}
\end{minipage}%
\begin{minipage}{.47\textwidth}
  \centering
  \includegraphics[height = 8cm,width = 6cm]{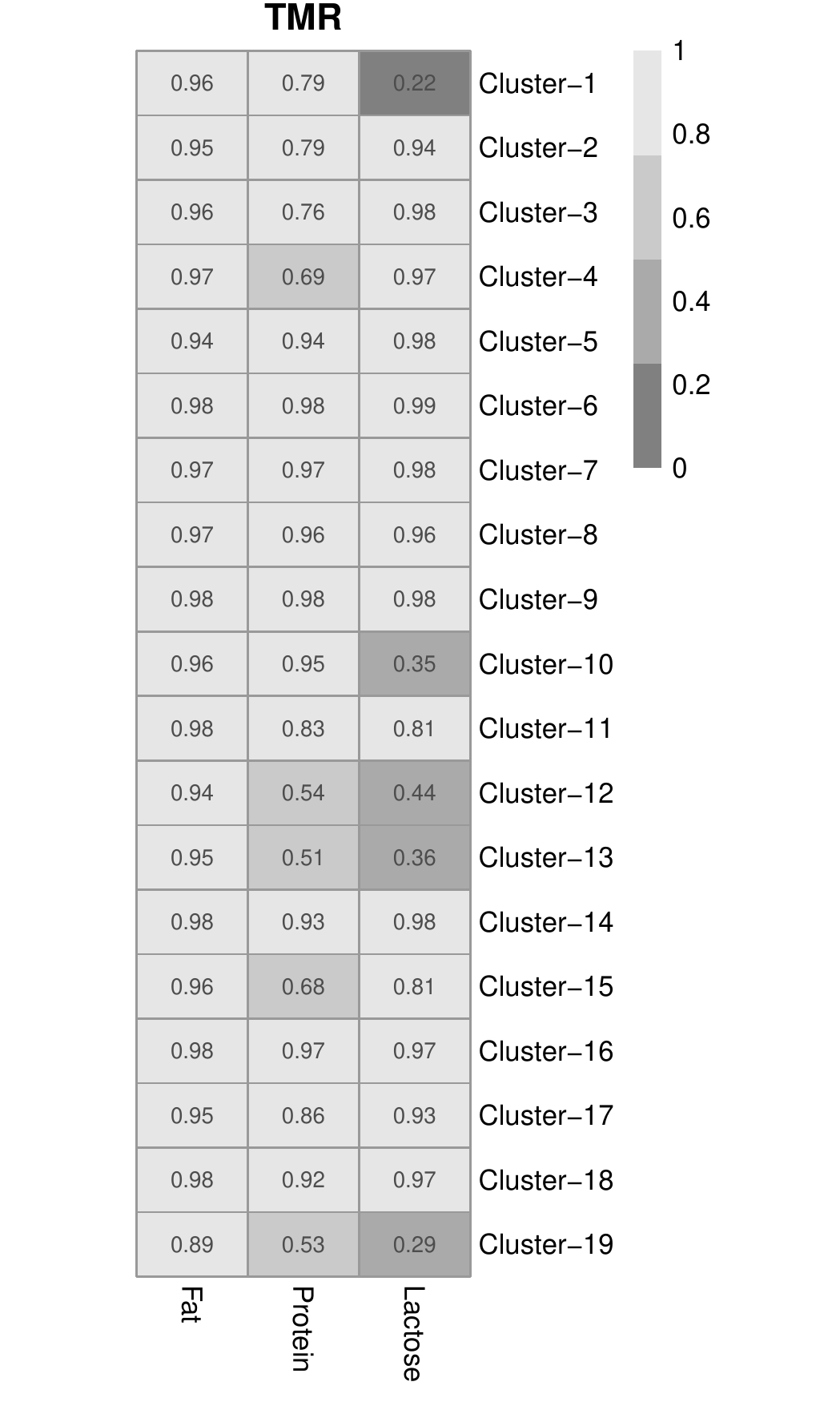}
\end{minipage}
  \caption{Adjusted R-squared values of the regression models where only the wavelengths in a specific cluster are used to predict the content of three different milk traits, namely Fat, Protein and Lactose contents. On the left the results for the pasture samples, on the right the ones for the TMR ones.}
 \label{fig:adjRsquared}
\end{figure}

In order to gain some further useful and practical knowledge about the phenomenon that we are studying, we considered a cluster-specific predictive analysis. Therefore, we run different linear regression models, separately for pasture and TMR samples, where the covariates are given by the spectral measurements at the wavelengths belonging to a cluster, while the response variables are the content of fat, protein and lactose in the samples. These analyses, as briefly mentioned in the introduction, are important in order to understand if spectroscopy data may be fruitfully used in order to predict some important features of the milk in a rapid and non-expensive way. Moreover, in our case, the predictions are based only on a small subset of variables, namely those assigned to a specific cluster, thus alleviating high-dimensionality induced issues. In Figure \ref{fig:adjRsquared} we report the results we obtained in terms of the \emph{Adjusted R-squared} index. At first glance it seems that the content of fat in the milk samples is easy to predict, regardless of the specific spectral region considered and of the diet regimens. Moreover, the predictive performances are generally higher for the TMR samples in comparison to the pasture ones. A more careful examination of the results, if paired with the suggestions obtained about the wavelengths clustering structures, allows us to give some further indication about how the information carried in some spectral regions is different depending on the diet. For example, from Table \ref{tab:table_clust_results} we can see how TMR cluster 15 find its correspondence with pasture clusters 16, 17 and 19. It is straightforward to see how, in terms of the \emph{Adjusted R-squared}, the wavelengths in the corresponding regions seem to produce better predictions of the lactose content for the TMR than for the pasture milk samples. Similar indications can be found by carefully studying jointly the results shown in Table \ref{tab:table_clust_results} and Figure \ref{fig:adjRsquared}. 

The clustering results obtained, if paired with previously conducted studies, can lead to other relevant insights. For example, the work by \citet{picque1993monitoring} suggests that the measurements in the region spanning from 1515cm$^{-1}$ to 1593cm$^{-1}$ are characteristic of the lactate ion. On the other hand, the regions from 1040cm$^{-1}$ to 1100cm$^{-1}$ and from 1298cm$^{-1}$ to 1470cm$^{-1}$ are related to galactose component of milk. Note that, from a practical standpoint, both lactate and galactose can be seen as indicators of the milk quality. As is visually clear in Figure \ref{fig:cluster_waves}, the wavelengths pertaining to the lactate ion region mainly belong to pasture cluster 8 and to TMR cluster 7; a closer inspection of Table \ref{tab:table_clust_results} reveals how these groups are strongly related in the two partitions, giving an additional indication about the coherency of the clustering results. On the other hand the wavelengths belonging to the galactose regions are split into groups 2, 5 and 7, for the pasture samples, while they are mainly associated with groups 3 and 6 for the TMR samples; again with a strong correspondence among these clusters visible in the confusion matrix. Moreover, the results obtained from the cluster specific regression analyses show how these groups are among the best ones for predicting the lactose content in the milk samples. Since lactose is a disaccharide molecule formed by the linking of the galactose and glucose monosaccharide modules, these results serve as a confirmation of the practical utility of the variable partitions obtained from the model.

\begin{figure}[!t]
\centering
\begin{minipage}{.47\textwidth}
  \centering
  \includegraphics[height = 6cm, width=6cm]{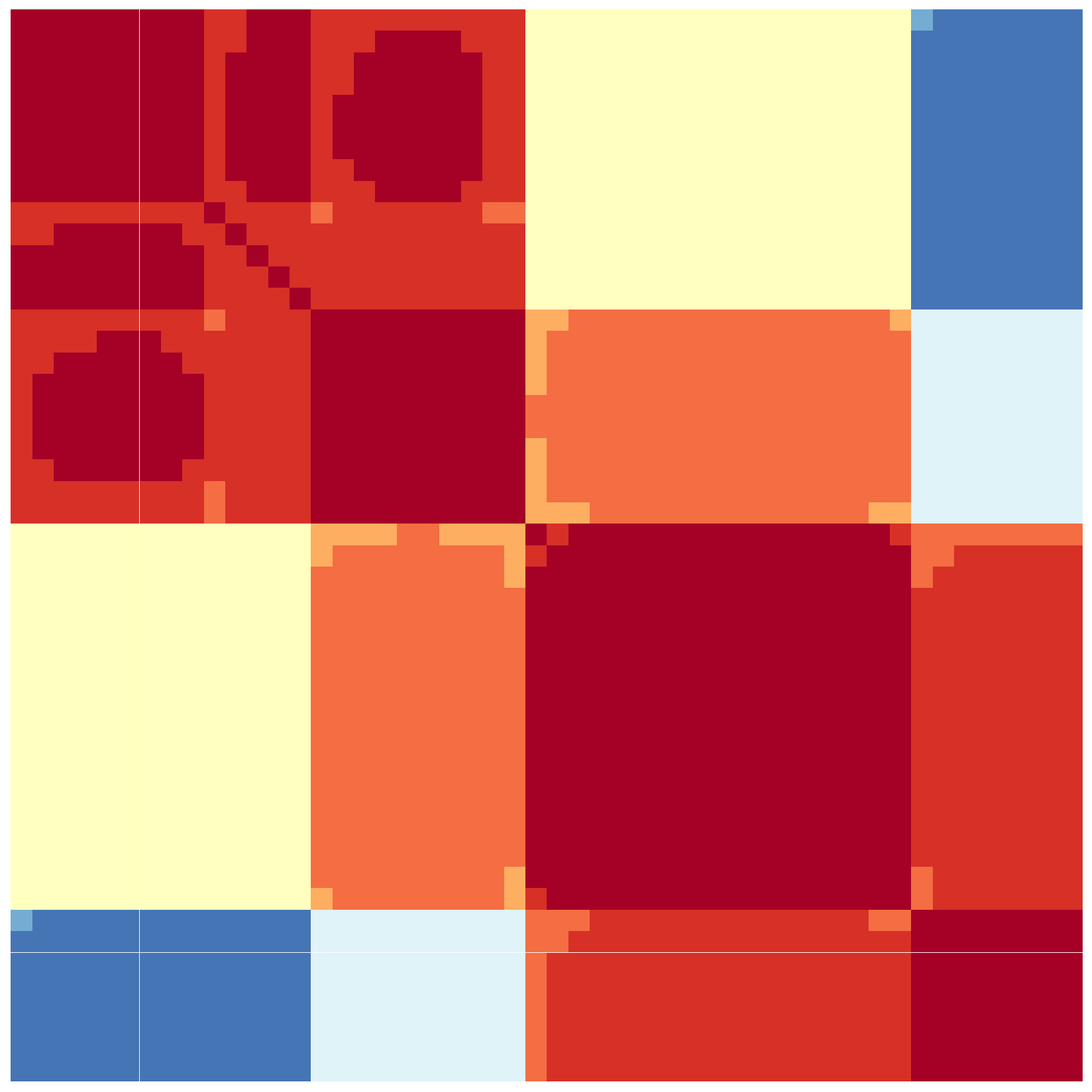}
\end{minipage}%
\begin{minipage}{.47\textwidth}
  \centering
  \includegraphics[height = 6cm,width = 6.5cm]{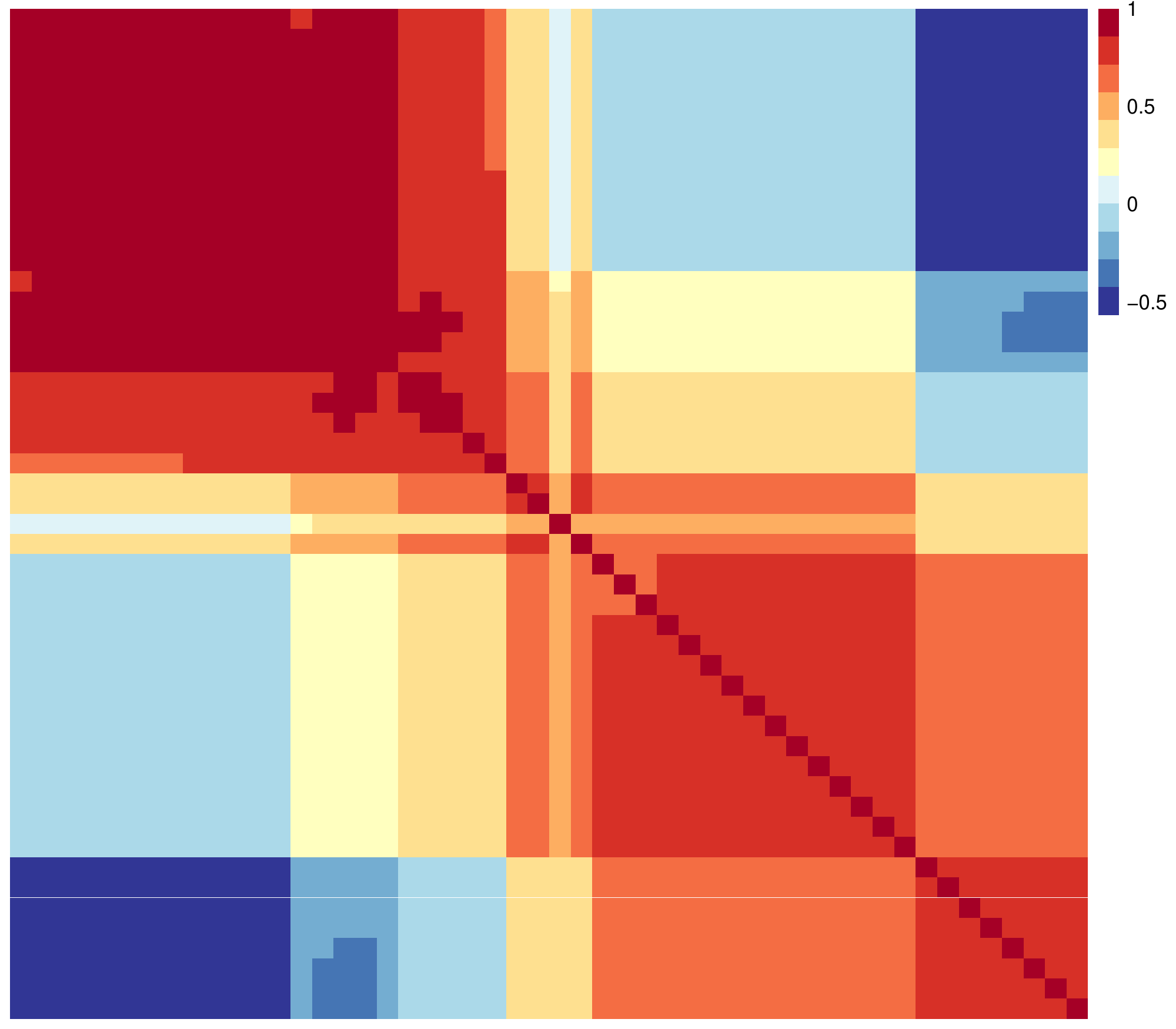}
\end{minipage}
 \caption{Estimated correlation sub-matrices corresponding to the first 50 wavelengths computed on the milk samples produced by pasture fed cows (on the left) and TMR fed cows (on the right).}
 \label{fig:diffwaves}
\end{figure}

Lastly, note that the correlation matrices estimated by using the proposed methodology can be used as an exploratory tool to deepen our knowledge about the relations among wavelengths in a specific spectral region. For example, if we focus our attention on the correlations among the first 50 wavelengths. From a visual inspection of Figure \ref{fig:estCor}, the wavelengths in this region seems to relate one to the other differently depending on the diet; this region is shown more closely in Figure~\ref{fig:diffwaves}. From Figure~\ref{fig:diffwaves}, we can see how the relationships among the spectral values at these wavelengths is highly dependent on the diet in this region. If we compare the two sub-matrices using the metrics considered previously, we obtain that $\text{MSE}(\hat{R}_{\text{Pasture}}^{(1:50)},\hat{R}_{\text{TMR}}^{(1:50)}) = 0.058$ and $\text{RV}(\hat{R}_{\text{Pasture}}^{(1:50)},\hat{R}_{\text{TMR}}^{(1:50)}) = 0.931$, thus providing an indication of stronger discrepancies in this spectral region compared to the one witnessed among the full covariances. Again, considering jointly these indications with the results outlined above, we can hypothesize the reasons behind this difference. More specifically, in this case, the initial wavelengths seem to give consistently better performances when predicting the protein content for the TMR samples compared to the pasture samples. Similar analyses can be conducted also for other spectral regions, depending on the specific application interest. 

\section{Discussion and further work}\label{sec:finalremarks}
In this paper we have presented a modification of a standard Factor Analysis model where factor loadings matrix is reparameterised so that redundancy in the originally observed variables can be detected. Moreover, to estimate the proposed model, a flexible Metropolis within Gibbs sampler has been implemented. Our method yields a parsimonious representation of strongly dependent high-dimensional data with complex correlation structures. In fact, the specification we propose, entails a huge reduction of the number of parameters to be estimated with respect to a standard FA model. At the same time, as a direct consequence of the specification itself, the model yields a grouping of the original variables when mapping them into the lower-dimensional subspace. The subsequent partition throws light on the relations among the observed features and about their possible redundancies. These indications, when supported by subject matter knowledge, can be translated into practical knowledge about the phenomenon under study. 

Our proposal was directly motivated by an application to vibrational spectroscopy data analysis and showed good performances on the dairy feed experiment data under investigation, both in terms of correlation reconstruction and interpretability of the results. 

Spectroscopy data usually present some recurring challenges, from a statistical perspective, as they are high-dimensional with strong and peculiar correlation structures among the wavelengths, possibly entailing complex redundancies. The model we introduced has been proven particularly useful in the given context since it has provided a parsimonious characterization of the correlation matrix. From a practical point of view, this allowed to gain relevant knowledge and to highlight differences among different milk samples, thus possibly helping in authenticity assessment and in preventing food adulteration in the future. Moreover, by clustering the variables, it provided interesting insights about which spectral regions carry the same information implying that, even if possibly far from each other, they may be influenced by similar chemical processes. Lastly note that our proposal can be thought as a sort of starting point when building classification tools aiming to discriminate samples according to the relations occurring among the observed features.  

Note that, the proposed methodology has been applied to MIR spectroscopy data but, in principle, its use may be extended to other data sharing some characteristics with the ones considered in this work. A possible interesting extension and research direction consists in the exploration of different possibilities concerning $\pi(Z)$, the prior distribution for the allocation matrix. When accounting for peculiar correlation structures in the data, it can be interesting, as we briefly mentioned in Section \ref{sec:likelihood_prior}, to explore prior distributions incorporating information about specific relations and constraints, such spatial or temporal ones, for the variables to be clustered. Another aspect that is worth examining is concerned with the model selection. In Section \ref{sec:modelselection} we introduced an initialization strategy that provides good indications about reasonable values for the number of factors $K$ and clusters $G$. Nonetheless several different approaches may be adopted and a thorough exploration of different model selection tools may be beneficial. As briefly mentioned in Section \ref{sec:modelselection}, a possible extension consists in allowing $K$ to go toward infinity and in considering shrinkage priors on the factor loadings as proposed in \citet{bhattacharya2011sparse,murphy2018infinite}. This strategy, possibly fruitful even for the determination of $G$, allows to circumvent the issues related to the selection of $K$, by automating the choice of the active factors, i.e. the ones with non-negligible loading values, in characterizing the covariance structure. Finally, as we mentioned above, our proposal can be thought as a stepping stone when building new classification tools. A possible straightforward strategy, pointing in this direction, would consist in embedding the model we introduced in a Mixture of Factor Analysis \citep[MFA,][]{ghahramani1996algorithm} framework thus allowing to perform classification and clustering of high-dimensional data.

\begin{appendix}
\vspace{0.5cm}
\section*{Posterior conditional distributions}\label{appendix}
\noindent {\bf Full conditional for the factor scores $U$} \\
\noindent Let denote, as we did in the Section \ref{sec:model_formulation}, $\tilde{\Lambda} = Z\Lambda_c$. The full conditional (\ref{eq:fullcond_scores}) for the factor scores $U$ is obtained following standard FA results as follows 
\begin{eqnarray*}
\pi(U | \dots) &\propto& \mathcal{L}(X | \Lambda_c, \Psi, Z, U)\pi(U) \\
&\propto& \exp\left\{-\frac{1}{2}\text{tr}[\Psi^{-1}(X-U\Lambda^T)^T(X-U\Lambda^T) ]\right\} \exp\left\{ -\frac{1}{2} \text{tr}[U^TU] \right\}  \\
&\propto& \dots \\
&\propto& \exp\left\{-\frac{1}{2}\text{tr}[(\mathbb{I} + \Lambda^T\Psi^{-1}\Lambda)(U-\tilde{U})^T(U - \tilde{U})] \right\} \\
\end{eqnarray*}
Therefore U is distributed as a matrix-normal random variable, $U \sim \mathcal{MN}_{n,K}(\tilde{U}, \mathbb{I}_n, (\mathbb{I}_K + \Lambda^T\Psi^{-1}\Lambda)^{-1})$.
Focusing on a single row of the factor scores matrix we obtain  
  $$
  \pi(u_i | \dots) \sim \mathcal{N}_K((\mathbb{I} + \Lambda^T\Psi^{-1}\Lambda)^{-1}\Lambda^T \Psi^{-1}x_i, (\mathbb{I} + \Lambda^T\Psi^{-1}\Lambda)^{-1}) \;\; \text{for} \; i = 1, \dots, n
  $$
  so that 
  \begin{eqnarray*}
  \mu_u &=& (\mathbb{I} + \Lambda^T\Psi^{-1}\Lambda)^{-1}\Lambda^T \Psi^{-1}x_i \\
  \Sigma_u &=& (\mathbb{I} + \Lambda^T\Psi^{-1}\Lambda)^{-1}
  \end{eqnarray*}

\vspace{0.5cm}
\noindent {\bf Full conditional for the unique factor loadings matrix $\Lambda_c$} \\
Using several times the properties of the trace operator and of the vectorization, the full conditional (\ref{eq:fullcond_lambda}) is obtained as follows 
\begin{eqnarray*}
\pi(\Lambda_c | \dots) &\propto& \mathcal{L}(X | \Lambda_c, \Psi, Z, U)\pi(\Lambda_c) \\
  &\propto& \exp\left\{ -\frac{1}{2}\text{tr}[\Psi^{-1}(X-U\Lambda_c^TZ^T)^T(X-U\Lambda_c^TZ^T)] \right\}\exp\left\{ -\frac{1}{2} \text{tr}[\sigma_\lambda^{-2}\Lambda_c^T\Lambda_c] \right\} \\
  &=& \exp\left\{ - \frac{1}{2}\text{tr}[(X-U\Lambda_c^TZ^T)\Psi^{-1}(X-U\Lambda_c^TZ^T)^T] -\frac{1}{2}\sigma_\lambda^{-2}\text{tr}[\Lambda_c^T\Lambda_c] \right\} \\
  & = & \exp\left\{ - \frac{1}{2}\left[ \text{tr}(U\Lambda_c^TZ^T\Psi^{-1}Z\Lambda_c U^T) -2 \text{tr}(X\Psi^{-1}Z\Lambda_cU^T) + \sigma_\lambda^{-2}\text{tr}(\Lambda_c^T\Lambda_c) \right] \right\} \\
  & = & \exp\biggl\{ - \frac{1}{2}\bigl[ \text{tr}(U\Lambda_c^TZ^T\Psi^{-1}Z\Lambda_c U^T) -2\text{vec}(\Lambda_c)^T\text{vec}(Z^T\Psi^{-1}X^T U) + \\
  && \sigma_\lambda^{-2}\text{tr}(\Lambda_c^T\Lambda_c) \bigr] \biggr\} \\
  & = & \exp\biggl\{ - \frac{1}{2}\bigl[ \text{vec}(\Psi^{-1/2}Z\Lambda_cU^T)^T\text{vec}(\Psi^{-1/2}Z\Lambda_cU^T) + \sigma_\lambda^{-2}\text{vec}(\Lambda_c)^T\text{vec}(\Lambda_c) - \\
  && 2\text{vec}(\Lambda_c)^T\text{vec}(Z^T\Psi^{-1}X^T U) \bigl] \biggl\} \\
  & = & \exp\biggl\{ - \frac{1}{2}\bigl[ ((U \otimes \Psi^{-1/2})\text{vec}(\Lambda_c))^T(U \otimes \Psi^{-1/2})\text{vec}(\Lambda_c) + \\
  && \sigma_\lambda^{-2}\text{vec}(\Lambda_c)^T\text{vec}(\Lambda_c) -2\text{vec}(\Lambda_c)^T\text{vec}(Z^T\Psi^{-1}X^T U) \bigl] \biggl\} \\
  &\propto& \exp\biggl\{ - \frac{1}{2}\bigl[ \text{vec}(\Lambda_c)^T(U^TU \otimes Z^T\Psi^{-1}Z + \sigma_\lambda^{-2}\mathbb{I})\text{vec}(\Lambda_c) - \\ 
  && 2\text{vec}(\Lambda_c)^T\text{vec}(Z^T\Psi^{-1}X^T U) \bigl] \biggl\} \\
\end{eqnarray*}
which is proportional to a pdf of a multivariate Gaussian random variable. Therefore we have that 
$$
\text{vec}(\Lambda_c) \sim \mathcal{N}_{G\times K}(\mu_\Lambda, \Sigma_\Lambda)
$$
where 
\begin{eqnarray*}
  \mu_\Lambda &=& (U^TU \otimes Z^T\Psi^{-1}Z + \sigma^{-2}\mathbb{I})^{-1}\text{vec}(Z^T\Psi^{-1}X^T U) \\
  \Sigma_\Lambda &=& (U^TU \otimes Z^T\Psi^{-1}Z + \sigma^{-2}\mathbb{I})^{-1}
  \end{eqnarray*}

\vspace{0.5cm}
\noindent {\bf Full conditional for the uniquenesses $\psi_j$'s} \\
Let define $M = (X-U\Lambda_c^T Z^T)^T(X-U\Lambda_c^TZ^T)$ and $M_{jj}$ as the $(j,j)$-th element of the matrix $M$. The full conditional (\ref{eq:fullcond_uniqueness}) for the uniquenesses $\psi_j$ for $j=1,\dots,p$ are obtained as 
\begin{eqnarray*}
  \pi(\Psi | \dots) &\propto& \mathcal{L}(X | \Lambda_c, \Psi, Z, U)\pi(\Psi) \\
  & \propto & |\Psi|^{-n/2}\exp\left\{-\frac{1}{2}\text{tr}[\Psi^{-1}M]\right\} \prod_{j=1}^{p} \frac{\beta^{\alpha}_j}{\Gamma(\alpha)}\psi_j^{-(\alpha + 1)}\exp\left(-\frac{\beta_j}{\psi_j} \right)\\
  & \propto & \left(\prod_{j=1}^{p} \psi_j^{-n/2}\psi_j^{-(\alpha + 1)}\right) \exp\left\{ -\frac{1}{2} \text{tr}[\Psi^{-1}M] - \beta_j\sum_{j=1}^p \psi_j^{-1}\right\} \\
  & = & \left(\prod_{j=1}^{p} \psi_j^{-(\alpha + \frac{n}{2} +1)}\right)\exp\left\{-\frac{1}{2} \sum_{j=1}^p \psi_j^{-1}M_{jj} - \beta_j\sum_{j=1}^p \psi_j^{-1} \right\} \\
  & = & \left(\prod_{j=1}^{p} \psi_j^{-(\alpha + \frac{n}{2} +1)}\right)\exp\left\{-\sum_{j=1}^p \psi_j^{-1}\left(\frac{M_{jj}}{2} + \beta_j\right) \right\}
  \end{eqnarray*}
  Therefore we obtain that $(\psi_j | \dots) \sim \text{InvGamma}(\alpha + n/2, \beta_j + M_{jj}/2)$
  so that $\beta_j^* = \beta_j + M_{jj}/2$.

\end{appendix}

\section*{Acknowledgements}
This publication has emanated from research conducted with the financial support of Science Foundation Ireland (SFI) and the Department of Agriculture, Food and Marine on behalf of the Government of Ireland under grant number (16/RC/3835) and the SFI Insight Research Centre under grant number (SFI/12/RC/2289\_P2).

\bibliographystyle{apalike}
\bibliography{biblio.bib}

\end{document}